\documentclass[a4paper,12pt]{article}
\usepackage{amssymb,amsmath,mathbbol,mathrsfs}
\usepackage[usenames,dvipsnames]{color}
\usepackage{hyperref}
\usepackage{xcolor}
\usepackage{stmaryrd}
\usepackage{authblk}
\usepackage{framed}
\usepackage{empheq} 
\usepackage{slashed}
\usepackage{hyphenat}
\usepackage{cite}
\usepackage{apacite}
\usepackage{natbib}
\usepackage{dsfont}
\usepackage{chngcntr}
\usepackage{enumerate}  

\usepackage{marginnote}    

\usepackage[left=.8in,right=.8in,top=.8in,bottom=.8in]{geometry}                  

\usepackage{sectsty} 
\allsectionsfont{\sffamily\mdseries\upshape} 
\usepackage{tocloft}

\makeatletter
\renewenvironment{abstract}{%
    \if@twocolumn
      \section*{\abstractname}%
    \else 
      \begin{center}%
        {\bfseries\sffamily\abstractname\vspace{\z@}}
      \end{center}%
      \quotation
    \fi}
    {\if@twocolumn\else\endquotation\fi}
\makeatother

\numberwithin{equation}{section}
\hypersetup{
	colorlinks=true,         
	linkcolor=MidnightBlue,          
	citecolor=BrickRed,        
	urlcolor=MidnightBlue            
}

\newcommand{\be}{\begin{equation}}
\newcommand{\ee}{\end{equation}}

\renewcommand{\d}{{\mathrm{d}}}

\newcommand{\pp}{{\partial}}

\newcommand{\Diff}{{\mathrm{Diff}}}

\renewcommand{\bar}{\overline}

\renewcommand{\hat}{\widehat}

\newcommand{\RR}{\mathds{R}} 

\newcommand{\cint}{{\int\kern-.87em{<}}}
\newcommand{\sint}{{\int\kern-.75em{\sim}}}
\newcommand{\fint}{{\int\kern-1.00em{\int}}}

\let\oldmarginpar\marginpar
\renewcommand\marginpar[1]{\oldmarginpar{\color{red}\raggedright\footnotesize #1}}

\begin{document}

\title{The hole argument meets Noether\rq{}s theorem}
\author{Henrique Gomes\footnote{gomes.ha@gmail.com}\\
Oriel College, University of Oxford}

\maketitle

\begin{abstract}
The hole argument of general relativity threatens a radical and pernicious form of indeterminism. One natural response to the argument is that points belonging to different but isometric models should  always be identified, or \lq{}dragged-along\rq{}, by the diffeomorphism that relates them. In this paper, I first criticise this response and its construal of isometry: it stumbles on certain cases,  like Noether's second theorem. Then I go on to describe how the essential features of Einstein\rq{}s `point-coincidence\rq{}  response to the hole argument avoid the criticisms of the `drag-along response\rq{} and are compatible with Noether\rq{}s second theorem. 
\end{abstract}

\section{The hole argument and the drag-along response}
A  natural response to the indeterminism threatened by the hole argument has recently gotten a second-wind. This  is \lq{}the drag-along response\rq{}:  spacetime points belonging to different but isometric models should always be identified, or \lq{}dragged-along\rq{}, by the diffeomorphism that relates them. This response implies that the action of  the Lie derivative, if understood as the limit of a pull-back by a diffeomorphism, vanishes identically. Agreed, in defining the Lie derivative, one need not employ the \lq{}drag-along\rq{} construal of isometry. However, in some applications, such as in the proof of Noether\rq{}s second theorem, we need to understand Lie derivatives as limits of the pull-backs by the very same isometries that motivated the drag-along argument. In other words, isometries, as construed in the drag-along, lead to the wrong inferences in the context of Noether\rq{}s second theorem.  The conclusion is that the drag-along response cannot be correct for  all spacetime symmetries. And this limitation raises the question: is there a single interpretation of the theory and its isomorphisms that allows us to refute indeterminism \emph{and} apply the conclusions of Noether\rq{}s theorems?  Here I propose a way to answer `yes\rq{} to this question. 

Here is the prospectus for this paper: In Section \ref{sec:hole} I will (very) briefly summarise the hole argument; this Section is not meant as a thorough introduction to the hole argument: there are already many  sources for that (see e.g. \citep{Pooley_draft, GomesButterfield_hole1,  Pooley_Read, SEP_hole} and references therein). In Section \ref{sec:drag} I will similarly summarise the \lq{}drag-along\rq{} response to the hole argument. Then in Section \ref{sec:Noether}, I will similarly summarise Noether\rq{}s second theorem in the case of general relativity, emphasising its comparison of isomorphic models by a map that is not the isomorphism that relates them. Finally, in Section \ref{sec:resolve}, I will trace a path to Noether\rq{}s theorem that goes between the Scylla  of a  non-trivial  Lie derivative and the and Charybdis of the hole argument. 

\subsection{The hole argument, in brief}\label{sec:hole}

After being rediscovered by John Earman, John Norton, and John Stachel in the 1980\rq{}s (see e.g. \citep{EarmanNorton1987}), whether by consensus or resignation, philosophical debate about the hole argument enjoyed two decades of relative quietness, from the mid-90\rq{}s to the mid-2010\rq{}s (cf. \citep{Weatherall_pre} for a recent history and philosophical appraisal). But vigorous debate was reignited by \cite{Weatherall_hole}, who claimed that a natural, mathematical understanding of isomorphism sufficed as a response to the argument. After a brief introduction to the hole argument, I will similarly summarise the `drag-along\rq{} response, that was at the core of \cite{Weatherall_hole}\rq{}s claims. 

Thus, in the interest of brevity, let us  focus on the problem that \cite{Pooley_Read} call \lq{}the problem of indeterminism\rq{}, that most bothered Einstein in his struggles toward general relativity  (see \citep{Giovanelli2021} for a history of these struggles).  

 Given a solution $g_{ab}$ (a metric) of general relativity over the spacetime $M$, with initial data $\Delta$ on a Cauchy surface $\Sigma \subset M$ (a global instant), we can obtain infinitely many other solutions, with the same $\Delta$, threatening indeterminism.  The alternative solutions are obtained by smoothly 'reshuffling' the manifold points with any $d: M \rightarrow M$, a one-to-one map on spacetime with $d$ and $d^{-1}$ smooth. In more detail, since  $d$ (and $d^{-1}$) takes smooth curves to smooth curves, it will induce a map on tangent vectors and co-vectors---pointwise linear isomorphisms of the tangent and cotangent spaces---denoted by $d_*$ and $d^*$ respectively, and also on  their tensor products; so it induces such a map on the metric tensor. If we pick a $d$ such that  $d^*_{|_\Sigma} =\mathsf{Id}$, this map 'slides' the metric's profile of values, obtaining $d^* g_{ab}$, and leaving the values of $g_{ab}$ at $\Sigma$ untouched, and so preserving $\Delta$. Both the original metric and the metric with the profile of values \lq{}slid over\rq{} by the smooth reshuffling are solutions of the Einstein equations, since these equations covary under this reshuffling.\footnote{The Einstein equations are $R_{ab}(x)=GT_{ab}(x),\quad \forall x\in M$, where $R_{ab}$ is the Ricci tensor, $G$ is a gravitational constant in natural units, and $T_{ab}$ is the energy-momentum tensor of the other fields populating spacetime. Since the equations are valid for all $x\in M$, and $R_{ab}$ and $T_{ab}$ are tensors, sliding them over via the pull-back of a diffeomorphism will just shuffle the points around and so the new fields will still satisfy the Einstein equations. This same symmetry group also extends to the quantum regime; both the classical and the quantum symmetry arise from invariance of the Einstein-Hilbert action under diffeomorphisms, which will be how we conceive of the symmetries for the purpose of Noether\rq{}s theorem, in   Section \ref{sec:Noether}. `Covariance anomalies\rq{} occur only in somewhat exotic gravitational theories: they may only appear in parity-violating theories in 4k+2 dimensions, when Weyl fermions of spin 12 or 32,
 or self-dual antisymmetric tensor, fields are coupled to gravity. (For Dirac fermions there is no trouble.) See \citep{AlvarezWitten}.  \label{ftnt:EFEs}}

 Broadly, the question invoked by the hole argument is: how should we interpret the differences between $g_{ab}$ and $d^* g_{ab}$? Einstein struggled with this question in the final years before the birth of general relativity, coming finally to conclude that \lq\lq{}physical significance should only be attributed to point coincidences\rq{}\rq{}, of e.g. material worldlines (see \citep{Giovanelli2021}).\footnote{\lq\lq{}All our space-time verifications invariably amount to a determination of space-time coincidences. If, for example, events consisted merely in the motion of material points, then ultimately nothing would be observable but the meeting of two or more of these points.\rq\rq{}\citep{Einstein_points}.} 

\subsection{Drag-along}\label{sec:drag}

Take the vacuum theory of general relativity to be defined by models,  $\langle M, g_{ab}\rangle$, with $M$ a smooth manifold and $g_{ab}$ a semi-Riemannian metric which satisfies some equations of motion.\footnote{This description is typical in what is known as \emph{the semantic view of theories}, which would also require the specification of a  representation relation to `the world\rq{}, or to the target of the models, but I will leave that unspecified. See \cite{Lutz_syntax} for a critical appraisal and contrast to the idea of laws as propositions about the world (the syntactic approach).  } The symmetries of general relativity are isometries: a natural notion of isomorphism for   manifolds equipped with (semi)-Riemannian metrics. And these isomorphisms are generated by the diffeomorphisms of $M$, which are automorphisms of $M$, seen as a smooth manifold.  The challenge of indeterminism here concerns the representation of isometric models: viz. whether they necessarily represent the same physical state of a target system.\footnote{Note that, by having fixed a target system in the representation relation, I have fixed the representational context and am therefore can omit recent discussions about `representational capacities\rq{}; see e.g. \cite{Fletcher_hole, Pooley_Read}. } 

From a more metaphysical perspective, we can articulate the problem in terms already familiar from the discussions between Leibniz and Newton on the nature of space, as reported in the famous \emph{Leibniz-Clarke correspondence}. In these terms, the question is whether different but isometric spacetimes are impossible: i.e. whether the isometric spacetimes are really \emph{Leibniz-equivalent} \citep{EarmanNorton1987}, and therefore should be regarded as being the very same spacetime. We will start with this perspective.

 As I described in the previous Section, symmetries can be seen as generated by a `smooth sliding\rq{} over the points of $M$, or as due to a reshuffling of points; this is what paves the way for the \emph{anti-haecceitist}  response to the hole argument, quite common within the philosophical literature (cf. \citep{Pooley_draft} for a thorough overview). 

 This jargon can be quickly summarized: purely haecceitistic possibilities involve individuals being ``swapped'' or ``exchanged'' without any qualitative difference.
 Anti-haecceitists about
spacetime points thus deny that different spacetimes could instantiate the same distribution of qualitative
properties and relations and differ only over which
spacetime points play which qualitative roles.\footnote{Anti-haecceitism, for my purposes, is indiscernible from what has been called `anti-individualism\rq{}; which \citet{Kment}, for instance, aptly describes  as \begin{quote} \lq\lq{}The view could perhaps be stated by saying that, even fundamentally speaking, there are indeed individuals, but there are no fundamental facts about which individual any one of them is. Individuals are, as it were, mere anonymous loci of instantiation of qualitative properties and relations, nameless pegs on which we can hang these properties and that we can connect by these relations. They are individuals without individuality.\rq\rq{}
\end{quote}} 

In this view, spacetime points can  be individuated by  a description  using only  qualitative properties; they can only be individuated by their participation in the network of relations to other spacetime points. And it is important to note that the denial of primitive identity for spacetime points is compatible with the existence of a `bare\rq{}  spacetime manifold composed of `numerically' distinct spacetime points. That is, the numerical distinctness of spacetime points can be stated without reference to the points\rq{} `primitive identity\rq{} (i.e. without the use of free, singular terms denoting individual elements of the set), and indeed, there can be distinct points without distinct qualitative properties---so the view is not committed to the principle of identity of indiscernibles in this sense.  Thus
 anti-haeccetism is \emph{prima facie} compatible with a thin version of `substantivalism' (see e.g. \cite[Sec. 3.3]{Pooley_draft}). 
In the words of \citet[p. 20]{Hoefer_hole}, one of the first to articulate the position:
 “[p]rimitive identity is metaphysically
otiose, and not a necessary part of the concept of a substance”. 

Carrying over this doctrine to the mathematical, model theoretic framework, the proposal  that \citet[Sec. 3]{GomesButterfield_hole1} call {\em the drag-along response} to the hole argument is a
 popular way to  `wear our anti-haecceitism on our sleeves' so to speak and thereby to resist the idea that the distance, relations, etc. between points varies across models, thus rendering all tensors trivially invariant under isomorphisms.  This response  builds on \citet[p. 77]{Field_soph}'s observation that 
\begin{quote}
“individuation of objects across possible worlds” is sufficiently tied to their
qualitative characteristics so that if there is a unique 1-1 correspondence
between the space-time of world A and the space-time of world B that
preserves all geometric properties and relations (including geometric relations
among the regions, and occupancy properties like being occupied by
a round red object), then it makes no sense to suppose that identification of
space-time regions across these worlds goes via anything other than this
isomorphism. \end{quote} 

And here is how   \citet[p. 12]{Stachel_Iftime_short} understand the idea:
   \begin{quote}
[...] the points of the manifold are not individuated independently of
the metric field. This means that space-time points have no inherent chronogeometrical
or inertio-gravitational properties or relations that do not depend
on the presence of the metric tensor field. This implies that when we drag-along
the metric, we actually drag-along the physically individuating properties and
relations of the points. Thus, the pull-back metric does not differ physically from
the original one. It follows that the entire equivalence class of diffeomorphically-
related solutions to Einstein’s empty space-time field equations corresponds
to one inertio-gravitational field.
Put in other words, [...points] 
 lack haecceity as individualized points of that space-time (``events'') unless and until a particular metric field is specified.
\end{quote}
In my own words,  the proposal is that if we are given an isomorphism    $d$ that sends the properties at point $x$ in one model to a point $y$ in another model, where $d(x) = y$, then we should ``rebrand'' $y$ in the codomain model as ``really being'' $x$;  or ``replace $y$ with $x$''. In this case, whatever properties $x$ represents in the target system, so will $y$.  

More specifically for general relativity and diffeomorphisms: suppose that we have two  metrics  on $M$ that are isomorphic: $g_{ab}$ and $\tilde g_{ab}=d^*g_{ab}$. If we are to compare what they say about points or regions of $M$, in principle we can use any  diffeomorphism  $f\in \Diff(M)$, taken to relate  a value of the metric over the point in the domain with a value over the point in the image. The drag-along response enjoins us to use the diffeomorphism that gives rise to the isometry, namely $d$, so that the  {tensor} $g_{ab}$ should be compared with the   tensor $\tilde g_{ab}$ by using $d$, thus bringing the value of $\tilde g_{ab}(d(x))$ to compare with the value of $g_{ab}(x)$. And indeed, by the definition of $\tilde g_{ab}$, the two tensors seem  identical using this standard of comparison: 
\be\label{eq:compare} (d^*(\tilde g_{ab}))(x):=(d^*(\tilde g_{ab}(d(x)))=g_{ab}(x).\ee
 Thus I agree that isometry gives  the only appropriate standard of comparison of isomorphic models  that brings the physical content of spacetime  regions and points   into strict \emph{coincidence}. This argument provides a way to understand any tensor on $M$ as trivially diffeomorphism-invariant, and it realizes anti-haecceitism in a very concrete manner.
 
The advocate for the drag-along response can go further:  saying that, compared  using any other diffeomorphism, the tensors  will differ---for   $f\neq d^{-1}$, unlike \eqref{eq:compare}, we (generically) have $g_{ab}(x)\neq (f^*\tilde g_{ab})(x)$---and that these are physical differences.
 I agree that these are,  \emph{pointwise}, physical differences. For example,  compared using the identity on $M$, at the point $x$, $g_{ab}$ may be flat  whereas $\tilde g_{ab}$ is not.  
 This is in line with \citet[p. 336]{Weatherall_hole}:\begin{quote}
When we say that $\langle M, g_{ab}\rangle$ and $\langle M, \tilde g_{ab}\rangle$ are isometric spacetimes,
and thus that they have all of the same invariant, observable
structure, we are comparing them relative to  [the isometry...] if one only considers
[the isometry], no disagreement arises regarding the value of the metric at
any given point, since for any point $x\in M$, $g_{ab}(x) = \tilde g_{ab}(f (x))$
by construction. 
\end{quote}   

Let us more precisely verify the statements of the two previous paragraphs.  Let us call   $f\in \Diff(M)$ the base set map used for pointwise comparison of $g_{ab}$ and $\tilde g_{ab}$. Above $f$ was set either equal to the (inverse of the) map $d$---that gives rise to the isometry, $d^*$---or to the identity, $\mathsf{Id}$; here I am generalizing it to any diffeomorphism.  It is clear that if we demand that the comparison of the metrics through $f$ matches at every point:
\be g_{ab}(x)=(f^*\tilde g_{ab})(x)=((d\circ f)^*g_{ab})(x),\quad \forall x\in M,
\ee
then $d\circ f$ is an automorphism of $g_{ab}$. Assuming $g_{ab}$ is generic, it has only the identity as a (trivial) automorphism, and so $f=d^{-1}$.\footnote{I once thought that there was some conceptual gain obtained by thinking of every metric as inhabiting a different manifold: so, we think of $g_{ab}$ over $M$ and $h_{ab}$ over $N$, with $M$ and $N$ diffeomorphic. The idea then is that there is one diffeomorphism, $d$, now $d\in \Diff(M, N)$, such that $d^*h_{ab}=g_{ab}$. I no longer see any gain  in this manouver. For  one could still use any diffeomorphism (or rather, its inverse) $f\in \Diff(M, N)$, in order to compare $g_{ab}$ and $h_{ab}$, and only one such diffeomorphism would bring the metrics into coincidence. Moreover, given one such $f$, we can obtain the group of diffeomorphisms $\Diff(M, N)$ by composition of $f$ and $\Diff(M)$.    }

In one way,  \cite{Weatherall_hole} is right: given isometric models $\langle M, g_{ab}\rangle, \langle M, \tilde g_{ab}\rangle$,   among all of the diffeomorphisms $f$ used to compare these two metrics pointwise, it is only the drag-along, $f=d$, that will bring physical quantities into coincidence. And this view is entirely compatible with the metaphysical doctrine of anti-haecceitism.

\subsection{The drag-along: limitations.}

For practical purposes, the main limitation of the drag-along response is that general relativity, and our other spacetime theories, in some circumstances use means of identifying points \emph{other than} by drag-along. And they need to do so, on pain of trivialising important constructions: even elementary ones like the Lie derivative;  or more complex ones, like the   space of asymptotically  flat models. I will now explain only the first danger of trivialization.

 More precisely, about the Lie derivative:  for $f_t$ the flow of a vector field $X^a$, i.e. such that, $\forall p\in M, \quad \frac{d}{dt}(f_t(p))=X(p)$, the Lie derivative is usually defined using both the isometry induced by $f_t$ and the identity map for the base set $h=\mathsf{Id}$: we drag the tensors $f_t^*g_{ab}$ over the fixed base set $M$ (writing out the Lie derivative):
\be\label{eq:Lie}
{\mathcal{L}}_{{X}}{g}_{ab}(x):=\lim_{t\rightarrow0}\frac{1}{t}({g}_{ab}(x)-f_t^*{g}_{ab}(x)).
\ee
But if we instead use the drag-along to compare the metrics in the definition of the Lie derivative, we obtain: 
\be\label{eq:Lie_drag}
{\mathcal{L}}_{{X}}{g_{ab}}(x)=\lim_{t\rightarrow0}\frac{1}{t}({g_{ab}}(x)-f_t^*{g_{ab}}(f_t(x))\equiv 0 \; !
\ee Though a Lie derivative could be defined algebraically---as being a derivation satisfying certain axioms, such as commutation with the exterior derivative and with the contraction (or interior product)---this is not how it is mostly used. Indeed, by  restricting ourselves to such an algebraic definition we would lose the straightforward  relation between the Lie derivative and its flow, since that relation requires us to understand the pull-back along a diffeomorphism. 

But one need not resort to such an interpretation of the Lie derivative: we could just refrain from interpreting all diffeomorphisms as isomorphisms of the theory. The context of application of diffeomorphisms need not be an investigation of symmetries and there seems to be a consistent way to demarcate the application  of the Lie derivative when discussing symmetries from other applications. 

The deeper problem, as we will see in more detail in Section \ref{sec:Noether} below, is that the Lie derivative  allows us to obtain local conservation laws from Noether's second theorem. In this application, the Lie derivative is to be understood as arising from symmetries. So we cannot at the same time endorse a mandatory drag-along understanding of the symmetries of general relativity---the drag-along response to the hole argument---and the application of Lie derivative in Noether\rq{}s theorems about symmetries. Is there a coherent interpretation of the theory and its isomorphisms that allows us to rebuke indeterminism \emph{and} apply the conclusions of Noether\rq{}s theorems?

\section{Noether\rq{}s second theorem}\label{sec:Noether}

Here I will briefly state Noether\rq{}s second theorem, in a form that is useful for the main argument of this paper. For a more complete treatment, see \citep{Olver_book}.

We start by assuming that $\varphi$ is some field on spacetime, whose dynamically possible models are determined by  a Lagrangian scalar function $L(\varphi)\in C^\infty(M)$, as those that extremize the integral of this function over $M$ (called the \emph{action functional}), a condition which we write as:\footnote{The most geometric way to understand this equation is to think of $L$ as a scalar function on the space of models, and find the models where the functional gradient of this function vanishes.\label{ftnt:delta_L}} 
 \be\label{eq:delta_L} \delta \int_M L(\varphi)=0.\ee
 Equivalently, after successive integration by parts, we can write the conditions \eqref{eq:delta_L} as yielding equations of motion, up to boundary terms: 
 \be\label{eq:eom}
 \delta L=\mathsf{EL}(\varphi)_I\delta \varphi^I+\d \theta(\delta\varphi),
 \ee
 where we adopt summation conventions, $\mathsf{EL}(\varphi)_I$ is the Euler-Lagrange functional (the left-hand part of the Euler-Lagrange equations) which has one component for each component of $\delta \varphi$, denoted by the letter $I$;  and $\theta$ is a linear operator on variations of the fields, but it is a differential form of codimension one on spacetime (i.e. it is a boundary term). The Euler-Lagrange term is obtained from the variation of the Lagrangian along the field components by successive integration by parts, used in order to remove derivatives from the $\delta\varphi^I$; the remaining term is necessarily a total derivative, $\d\theta(\delta\varphi)$.\footnote{ 
 There are a couple of comments regarding the uniqueness of the several terms involved in \eqref{eq:eom} that we should address. First, for a fixed Lagrangian, the boundary term $\theta$   has an ambiguity: $\theta\rightarrow \theta+\d \kappa$, where $\kappa$ is an arbitrary form of codimension two on spacetime. Second, there may be more than one Lagrangian that yields the same Euler-Lagrange  part of the equations; the most common examples involve addition of terms to $L$ that don't depend on the fields (so that their variation vanishes), and additions that amount to a general shift of the boundary term $\theta\rightarrow \theta'$, which are hard to quantify. Here I will assume the Lagrangian is fixed up to boundary terms by further requirements that are left implicit (such as locality and, in the few cases in which that is not enough to eliminate unwanted alternatives,  the vaguer `simplicity' constraint). }
 
Suppose that, for any value of $\varphi$, there is a family of  transformations $\delta_\xi\varphi$, whose parameters $\xi$ form an algebra with commutator $[\bullet, \bullet]$,  such that $\delta_\xi\delta_{\xi'}\varphi-\delta_{\xi'}\delta_\xi\varphi=\delta_{[\xi,\xi']}\varphi$,\footnote{This condition is important in order to ensure that the symmetry-related values of the fields form an `orbit', or an integral submanifold in the space of models of the theory. The condition assumes that $\xi$ does not depend on $\varphi$: for the full explication of these equations in terms of the geometry of the space of models, and a generalisation to the case where $\xi$ is model-dependent, see \citep{GomesHopfRiello, GomesRiello_new}. } and so that $\delta_\xi L=0$. So from \eqref{eq:eom}, now omitting indices: 
\be\label{eq:xi_var} \delta_\xi L=\mathsf{EL}(\varphi)\cdot\delta_\xi \varphi+\d \theta_\xi=0,
 \ee

Very broadly, Noether's second theorem  follows from \eqref{eq:xi_var}. Here is a short proof: assume that the symmetries are malleable, or local, so that we can restrict to those $\xi$ such that $\xi_{|B}=0$.  This implies that $\d \theta_\xi{}_{|B}=0$.\footnote{The integration by parts required to get the variation in the form of \eqref{eq:xi_var} implies $\d\theta_\xi$ is linear on $\xi$; cf. \cite{Lee:1990nz}.}  Now,  there are  inner products $\langle\bullet, \bullet\rangle$ and $\langle\langle\bullet, \bullet\rangle\rangle$,  on the (tangent bundle of) the space of models and of the symmetry group, respectively, so that
\be\label{eq:gen_delta}\delta_\xi\int_M L(\varphi)=\int_M \mathsf{EL}(\varphi)\delta_\xi \varphi=\int_M\langle \mathsf{EL}(\varphi), \delta_\xi \varphi\rangle=\int_M\langle\langle \Delta^\dagger\mathsf{EL}(\varphi), \xi \rangle\rangle=0\ee
where $\Delta^\dagger$ is the formal adjoint of $\delta_\xi\varphi$, seen as a linear operator on $\xi$ (see \citep{fischermarsden} for a thorough, geometric formulatin of such inner products and adjoints in the space of fields of gauge theory and general relativity). Since  \eqref{eq:xi_var} must vanish for all such $\xi$, and $\xi$ can be chosen arbitrarily in different open subsets of $M$, we obtain a local equation that the Euler-Lagrange equations must satisfy everywhere, and which is valid off-shell:
\be\label{eq:Delta_dag}
\Delta^\dagger\mathsf{EL}(\varphi)=0.
\ee

To derive the required form of Noether\rq{}s second theorem in the more narrow application that we are focussing on---but still in a form broad enough to include any spacetime covariant theories, so that it would apply to higher-order corrections to classical general relativity coming from quantum gravity for example---we need two more requirements: 
 \begin{enumerate}[i)]
 \item \textbf{Lagrangian decoupling}:  We assume the field content of the models consists of a geometric part, contained in the metric $g_{\mu\nu}$, and the non-geometric parts, such as other force fields and their charged matter sources, jointly labelled by $\psi$. Then we assume that the Lagrangian decouples into one part that \emph{only} depends on the geometry: 
 \be\label{eq:decoup} L(\varphi)=L_{\text{\tiny{geom}}}(g)+L_{\text{\tiny{rest}}}(g,\psi).
 \ee
 \item \textbf{Minimal coupling:}  The energy-momentum tensor density can be obtained, up to boundary terms, by: 
 \be\label{eq:current} T_{\mu\nu}\hat{=}\frac{\delta L_{\text{\tiny{rest}}}(g, \psi)}{\delta g_{\mu\nu}}, 
 \ee 
where $\hat{=}$ denotes equality up to boundary terms. 
 If  equation \eqref{eq:delta_L} is like a gradient,  equation \eqref{eq:current} is like a directional derivative in the space of models, taken along directions that  vary only the metric, leaving the other fields fixed.  So the energy-momentum tensor is the sensitivity of the non-geometric part of the Lagrangian to the background metric. 
 \end{enumerate}

We focus on the case where the infinitesimal symmetries are generated by  vector fields, i.e. $\xi\rightarrow X\in \mathfrak{X}$ (the infinitesimal generator of a diffeomorphism), so that the Lie-algebra commutator is just the commutator on vector fields and:
\be \delta_\xi \varphi\rightarrow \delta_X g_{\mu\nu}:=\mathcal{L}_X g_{\mu\nu}=\nabla_{(\mu}X_{\nu)},
\ee 
where we used  \eqref{eq:Lie} and the Levi-Civita connection in the last equality, and round brackets denote anti-symmetrization of indices.\footnote{Of course, the Lie derivative obeys the required property of an infinitesimal symmetry generator: 
\be \mathcal{L}_X\mathcal{L}_{X'}g_{\mu\nu}-\mathcal{L}_{X'}\mathcal{L}_Xg_{\mu\nu}=\mathcal{L}_{[X,X']}g_{\mu\nu}.
\ee}

Putting this back on \eqref{eq:gen_delta}, we get, for variations purely along the geometric directions, assuming the vector field $X$  vanishes on the boundary of $M$: 
\be\label{eq:2_var} 0=\frac{\delta}{\delta g_{\mu\nu}}\left(\int_M L\right) \cdot \mathcal{L}_X g_{\mu\nu}=\int_M\left(\mathsf{EL}(g)_{\mu\nu}+T_{\mu\nu}\right)\mathcal{L}_Xg_{\mu\nu}=\int_M\left( \mathsf{EL}(g)_{\mu\nu}+T_{\mu\nu}\right)\nabla_{(\mu}X_{\nu)},
\ee
where $\cdot$ in the first term on the right-hand-side denotes contraction of a field-space one-form with a vector on field-space: we replace the appearance of $\delta g_{\mu\nu}$ on the result of the variation by $\mathcal{L}_Xg_{\mu\nu}$.  And, as in \eqref{eq:current}, from we get  $\mathsf{EL}(g)_{\mu\nu}$ from integration by parts  of $\frac{\delta L_{\text{\tiny{geom}}}(g)}{\delta g_{\mu\nu}}$ (removing derivatives from $\delta g_{\mu\nu}$). 

Since equation \eqref{eq:2_var} needs to hold even for kinematically possible models that do not obey the dynamical equations, i.e. we have used only invariance of the action, and not  the vanishing of the Euler-Lagrange functions, we get, after integration by parts, the functionally independent sets of equations: 
 \be \nabla^\mu T_{\mu\nu}=0\,\quad \text{and}\quad \nabla^\mu\mathsf{EL}(g)_{\mu\nu}=0.
 \ee
 The second set, of geometric equations, are implied by geometric constraints, coming from the Bianchi identities. For instance, for the particular choice of the Einstein-Hilbert Lagrangian, $L_{\text{\tiny{geom}}}(g)=R\sqrt{g}$, which is the Ricci scalar volume density, whose variation, up to boundary terms, gives the Einstein tensor, $G_{\mu\nu}$, and that is  covariantly divergenceless due to the contracted Bianchi identity. 
 
 On the other hand, for the first set of equations, each choice of matter fields gives its respective energy-momentum tensor density. So Noether\rq{}s second theorem ensures conservation of the energy momentum tensor of any field that satisfies the conditions (i) and (ii) (of Lagrangian decoupling and minimal coupling).  It is a non-trivial entailment that is unavailable with a drag-along understanding of isomorphisms. 

\section{Resolving the tension}\label{sec:resolve}

\subsection{Clearing the ground}
I should make it clear that I agree with \cite{Mundy1992, Weatherall_hole, BradleyWeatherall_hole} in many respects: mathematically,  the agreed-upon notion of isomorphisms for pseudo-Riemannian manifolds is isometry, and for that reason, isometric manifolds are, for what concerns any substantive mathematical   theorem (e.g. that you could find in a textbook, for instance, \citep{Oneill}), the same. 

 \cite{Weatherall_hole, BradleyWeatherall_hole}   associate  drag-along idea  with  \cite{Mundy1992}, who distinguishes a theory's synthetic language---that is only able to expresses qualitative, non-singular facts---from a theory's metalanguage, in which we are able to talk about points singularly, i.e. without a definite description using qualitative properties (see \cite[Sec. 3.2a]{Samediff_1a}). 
 Isomorphisms --- expressed in the metalanguage---will map objects singled out by the same description into each other.

However, physicists need an object language where the difference between isometric Lorentzian manifolds is expressible,
even if for purely formal purposes: any textbook of general relativity will work directly with tensor fields on a manifold, and not with an object that corresponds to their isomorphism-equivalence class.  As argued in \citep{Cudek_hole},  there is no non-set-theoretic foundation for mathematics that can completely handle the flexibility with which mathematical physicists use mathematical objects in the context of general relativity. Moreover, even if there were one, there is no prima facie reason to treat it as better than the set-theoretic foundation. 

Thus I take  \citet[p. 6]{BradleyWeatherall_hole}\rq{}s assertion that ``mathematicians generally intend to attribute to mathematical
objects only structure that is preserved by the relevant notion of isomorphism'' with a grain of salt. Mathematicians, like physicists, are opportunists: sometimes they use properties of mathematical objects that are not preserved by the relevant notion of isomorphism. 

And although  I believe that the drag-along is an important tool for assessing whether two models   represent the same physical possibility in general relativity, it is not the only tool, as I will discuss in Section \ref{sec:det}.\footnote{Indulging in a bit of metaphysical speculation, offered without further argument, I would say that what \emph{grounds} the existence of an isomorphism between models  is the structural similarity of two (isomorphic) mathematical objects---an isomorphism exists \emph{because} the two objects have the same structure---and not the other way around: they are not structurally the same \emph{because} there is such a map. Of course, nothing in my argument hangs on this metaphysical speculation.}


\subsection{Constructing diffeomorphisms that do not induce isometries}

Here I will describe diffeomorphisms that necessarily relate physically distinct points and yet preserve the Einstein-Hilbert action and also solutionhood of the Einstein equations. These diffeomorphisms cannot be interpreted via the drag-along, and they can be used in the proof of Noether\rq{}s second theorem; they are what \citet[Sec 3.1]{GomesButterfield_hole1} call a non-drag-along \lq{}threading\rq{} of spacetime points belonging to different models.  

The construction will proceed in two steps. First, in Section \ref{sec:active-passive}, I will show that given a chart, to each passive diffeomorphism, conceived of as a change of coordinates, we can unambiguously associate---both mathematically and conceptually---an active diffeomorphism: the kind of isomorphism of the smooth manifold we have been discussing. Second, in Section \ref{sec:Komar} I will construct a coordinate system from values of the metric, so that coordinates acquire  physical significance. Finally, by combining the arguments of the two Sections, I will show that one can associate a  change of these kinds of coordinates to maps between points with different physical properties and thereby build  diffeomorphisms that cannot lead to a `drag-along\rq{} comparison between models.

\subsubsection{The active-passive correspondence}\label{sec:active-passive}

 For a given $U\subset M$ as the domain of a chart $\phi: U\rightarrow \RR^k$, and  given an active diffeomorphism $f\in\text{Diff}(U)$, we can find a passive diffeomorphism 
\be\bar f:=\phi\circ f\circ \phi^{-1} \in\text{Diff}(\RR^k).\ee
And, conversely, given $\bar f\in \text{Diff}(\RR^k)$, we can find an active diffeomorphism
  \be\label{eq:passive_active} f:=\phi^{-1}\circ \bar f\circ \phi \in\text{Diff}(U).\ee
This diffeomorphism should be interpreted as follows: given coordinates $x$ for a point $p$, i.e. $\phi(p)=x$, we can, under a different chart, map $x$ to a different point, $q$, such that $\phi\rq{}(q)=x$. Different coordinate charts will ascribe a different value, a different spacetime point, to the same coordinates; so, having chosen a chart,  a change of coordinates gives rise to a unique active diffeomorphism acting  on the spacetime points.

 Agreed, we could have $f(U)\cap U=\emptyset$, in which case we clearly would not be able to register this diffeomorphism using solely a coordinate transformation in the domain of the chart $\phi$.  Having said that, if we restrict the diffeomorphisms to be connected to the identity, we need to only consider their generators, which are the infinitesimal flow of vector fields. And though these vector fields may be non-trivial at the boundary of the chart, they can still be represented within the charts, and thus at their intersection as well. Thus, given any atlas for the manifold---any covering of the manifold by a finite number of charts---we can patch together any infinitesimal active diffeomorphism using the infinitesimal passive diffeomorphisms in each chart of the atlas, and, by integration, recover the 1-1 correspondence.

\subsubsection{Physical coordinate systems}\label{sec:Komar}
In Section \ref{sec:drag}, we saw that, given some model,  the drag-along response avoids the indeterminism threatened by the multitude of isomorphic models. But the response did not sit well with standard mathematical tools, such as the Lie derivative, which require some relative shift between isomorphic metrics with respect to a fixed ``underlying'' set of  spacetime points.  But now I want to argue that  the Lie derivative can be  constructed from diffeomorphisms who (or whose pull-back) are not the isometries, and yet these diffeomorphisms can be represented as infinitesimal changes between physical coordinate systems (called \emph{representational conventions} in \citep{Samediff_1b, GomesButterfield_hole2, Kabel_et_al2024}). Let us see how this goes, and what it implies for determinism.

Let us for concreteness take the \emph{Kretschmann-Komar  coordinate system} \citep{Kretschmann_inv, Komar_inv}. The idea is to take the four scalar functions $\mathcal{R}^{(\mu)}(g_{ab}(x)),\,\, \mu=1, \cdots 4$, formed by certain real scalar functions of the Riemman tensor.\footnote{\cite{Komar_inv} finds these real scalars through an eigenvalue problem: 
 $$(R_{{a}{b}{c}{d}}-(g_{{a}{c}}g_{{b}{d}}+g_{{a}{d}}g_{{b}{c}}))V^{{c}{d}}=0,$$
 where $V^{{c}{d}}$ is an anti-symmetric tensor. } The use of an index in parenthesis emphasizes that this is just a list,  not the  components of a vector field.  For generic spacetimes (i.e. excluding Pirani's type II and III spaces of pure
radiation, in addition to excluding symmetric type I
spacetimes), these scalars are functionally independent. 

In such spacetimes, we \emph{define} coordinates  (omitting the dependence on the metric on the left-hand-side and the index's parenthesis):
\be\label{eq:Komar_coords} \bar x^\mu:=\mathcal{R}^\mu(g_{ab}(x)).\ee
In a less coordinate-centric language, the idea here is  to individuate spacetime points through \emph{some} of their qualitative properties, quantifying existentially over the points of $M$. That is, we define point $\bar x(g_{ab})$ through an inverse relation, in which $g_{ab}$ is the variable argument: e.g.  `$\bar x(g)$ is the point in which a given list of curvature scalars $\mathcal{R}^{(\mu)}(g)$, $\mu=1, \cdots, 4$ takes a specific list of values, $(a_1, \cdots, a_4)$'. This furnishes a qualitative individuation of \emph{spacetime points} across isomorphic and non-isomorphic models.\footnote{ \citet[p. 468]{Curiel2018} construes qualitative individuation of points similarly: \begin{quote}
\lq\lq{}Once one has the identification of spacetime
points with equivalence classes of values of scalar fields, one can as easily say
that the points are the objects with primitive ontological significance, and the
physical systems are defined by the values of fields at those points, those values
being attributes of their associated points only per accidens.\rq\rq{}
\end{quote}  }

Now, given some metric tensor $g^{\kappa\gamma}$ in coordinates $x^\kappa$, we can compute the metric in the new, $\bar x^\mu$ coordinate system as:
\be\label{eq:gbar} \bar g^{\mu\nu}=\frac{\pp \mathcal{R}^\mu}{\pp x^\kappa}\frac{\pp \mathcal{R}^\nu}{\pp x^\gamma}g^{\kappa\gamma}.
\ee
But this is just a family of 10  scalar functions indexed by $\mu$ and $\nu$. In the words of \citet[p. 1183]{Komar_inv}:\begin{quote} 
[it is] component by component a well defined scalar constructed from the metric tensor and its derivatives. If
we consider two metric tensor fields and ask whether
they represent the same physical situation, differing
perhaps by being viewed in different coordinate systems,
we now have a ready criterion for determining the
answer. Clearly, at corresponding points in any identifiation of the two spaces, the values of all scalars must
agree if the spaces are to be equivalent. We are therefore
compelled to identify points in the two spaces which
have the same ``intrinsic" coordinates [defined by \eqref{eq:Komar_coords}]. Furthermore at these corresponding points it is
necessary that the ten scalars [i.e. $ \bar g^{\mu\nu}$]
have the same values in the two spaces. 
Thus we find that the functional form of the 10 scalars
 [i.e. $ \bar g^{\mu\nu}$] as functions of the four scalars [i.e. $\mathcal{R}^\mu$]: (a) is uniquely
determined by the metric space independently of any
choice of coordinate system, and furthermore (b)
uniquely characterizes the space. \end{quote}

Thus spacetime points with different Komar coordinates necessarily have different physical properties. 
More straightforwardly about the Lie derivative: we can just track how quantities change from the point identified by e.g. $(1,2,3,4)_{\bar x}$ to an infinitesimally nearby point $(1+\delta \bar x^1,2,3,4)_{\bar x}$: this generates the Lie derivative of those quantities along $\frac{\partial}{\partial \bar x^1}$. The appearance of the Lie-derivative in Noether\rq{}s second theorem can therefore be explicitly understood through  this argument.\footnote{To be more explicit about the coordinate changes: we could easily have chosen a different set of scalars $\mathcal{R}^{(\mu)}$ provided we preserved their functional independence. For example, we could have multiplied $\mathcal{R}^{(1)}$ by two and taken the log of $\mathcal{R}^{(2)}$. Or we could take different linear combinations of $\mathcal{R}^{(\mu)}$; and  so on. Indeed, if we take $\mathcal{R}^{(\mu)}$ as coordinates of $\RR^4$, we could find an alternate set by applying any diffeomorphism (seen as a recombination of the original functions of the metric). Suppose we call two alternative sets of basis  in \eqref{eq:Komar_coords} $\bar x^{\mu}$ and $\tilde x^{\mu}$.  Now  within the  $\bar x^{\mu}$ choice, we can compare the point labeled by e.g. $(1,2,3,4)_{\bar x}$   to the point $(1,2,3,4)_{\tilde x}$. 
That is, as translated to the $\bar x^{\mu}$ choice, we can explicitly write the comparison of the point labeled by e.g. $(1,2,3,4)_{\bar x}$   to the point $(1,2,3,4)_{\tilde x}$; leaving the $\bar x^{\mu}$ subscript implicit:
\be (1,2,3,4)-(\bar x^1((1,2,3,4)_{\tilde x}), \bar x^2((1,2,3,4)_{\tilde x}),\bar x^3((1,2,3,4)_{\tilde x}),\bar x^4((1,2,3,4)_{\tilde x})).\ee This is the type of comparison, that, when pushed to the infinitesimal limit, generates the Lie derivative, which is here seen  as an infinitesimal version of a diffeomorphism that is in its turn seen as a change of the physical coordinate system.}

Incidentally, if $g_{ab}$ admits non-trivial automorphisms---or, in the standard nomenclature, admits Killing vector fields---then the $\bar x^{\mu}$ will not be a coordinate chart in the usual sense. For instance, if the metric is homogenous along some direction, we could have e.g.  $\bar x^1$  admitting a single value. In this case, we would automatically find that Lie derivatives along $\bar x^1$ vanish, for any function, which is what one would have expected in any case. If we want to have functions which are not homogeneous along that direction, and whose Lie derivative would not vanish, we would have to include more fields in the model, apart from the metric, and these could also be used to define physical coordinates $\bar x^{\mu}$. So the idea hangs together. 

 Importantly, note that, in this formulation, the Lie derivative does not require some non-qualitative identity of spacetime points: it only requires a continuous parametrization of spacetime points by their qualitative properties, in a manner compatible with anti-haecceitism.

Thus, using physical coordinate systems, through the passive-active correspondence of \eqref{eq:passive_active}, given an atlas (i.e. a finite collection of charts whose union covers $M$), any vector field in $\RR^4$ can be lifted to a vector field that is the infinitesimal generator of an active diffeomorphism: an active diffeomorphism that cannot be understood as a mapping between points with the same qualitative profile. 

To be explicit: such diffeomorphisms cannot be understood via the drag-along, and yet they generate a dynamical symmetry of the theory. That is, using physical coordinates such as  $\bar x$, we can write the Einstein equations (cf. footnote \ref{ftnt:EFEs}) using the coordinate metric \eqref{eq:gbar}, and changes of coordinates will leave these equations invariant.  Assuming the Einstein equations are satisfied at every spacetime point, the active transformations corresponding to a change of physical coordinate system  will preserve solutionhood of the equations because they map tensors that satisfy the Einstein equations at a point to tensors that satisfy it at another.  But had the Einstein equations been satisfied at some spacetime points but not others, such a transformation would \emph{not} be a symmetry. In other words,  a symmetry transformation  need not relate points with the same physical profiles; in order for a transformation to be a symmetry it is sufficient that it takes values of the fields that satisfy the equations of motion on one point, to values of the fields satisfying the equations of motion in another point. We have eliminated isomorphisms from the description of the metric, but a transformation can still be understood as a symmetry via a `team effort\rq{} of all the values of the fields over the manifold.

\section{Back to indeterminism?}

In the previous Section I showed that the Lie derivative can be defined along directions of local \lq{}physical changes\rq{}. The implication is that using the Lie derivative to study symmetries is compatible with an anti-haecceitist understanding of the spacetime metric. Thus, although  the  corresponding transformations preserve solutionhood of the equations of motion (or the value of the Einstein-Hilbert action functional) and are thus a dynamical or variational symmetry, they pose no threat of pernicious indeterminism.

In Section \ref{sec:det} I will  make this implication more precise, showing, as per conventional wisdom, that a description using physical coordinates  doesn\rq{}t allow indeterminism. This Section will argue that we can still  answer the threat of indeterminism with an anti-haecceitist understanding of spacetime, while preserving our hard-won conclusions about the physical Lie derivative\rq{}s use in the treatment of symmetries.  In Section \ref{sec:comp} I will describe a degenerate case, in which \lq{}physical\rq{} coordinates that are highly idealised still allow a kind of indeterminism about the metric. This second case can be seen as exploiting a lacuna in the definition of \emph{partial and complete} observables (cf. \cite{Rovelli_partial}). 

\subsection{Avoiding indeterminism}\label{sec:det}

\emph{ Within a single} representational convention, or choice of physical coordinate system, invariance under active diffeomorphisms is guaranteed, as  noted by Komar (cf. Section \ref{sec:Komar}). In the case of Komar variables the invariance directly follows from the definition of the coordinates. But more generally, a simple proof of this invariance is straightforward, and relies on the relationship between  physical coordinate systems and choices of gauge-fixing. In this Section, we will see how the general proof goes.

 Much as  we will do here, Komar sees gauge-fixing procedures as related to the definition of the physical coordinates through the values of curvature scalars, \eqref{eq:Komar_coords}. To see the relation explicitly, instead of defining new coordinates through \eqref{eq:Komar_coords}, we write out $g_{ab}$ in coordinates and, unlike \eqref{eq:Komar_coords}, have those same coordinates appear on the left and right hand side  of the equation, namely: 
\be\label{eq:point_gf} x^\mu=\mathcal{R}^{(\mu)}(g_{\kappa\gamma}(x)), \quad\text{or equivalently}\quad \pp_\nu \mathcal{R}^{(\mu)}(g_{\kappa\gamma}(x))=\delta_\nu^\mu,\ee  an equation that is to be solved for appropriate coordinates $x^\mu$, seen  as  functions of the metric.
But   \citet[p.1186]{Komar_inv} rightly highlights the difference between a physical system of coordinates and the idea that we are somehow `breaking [gauge] covariance':
\begin{quote}
The usual
argument, that employing coordinate conditions may
destroy the general covariance [...] does not apply in this case. For considering \eqref{eq:point_gf}
as a coordinate condition [as opposed to the definition of 10 scalar functions of the metric] is only a heuristic device to
make it easier to visualize how to manipulate the quantities with which we are dealing. In point of fact,
we know how to interpret these quantities as true
observables. (An apt analog in electromagnetic theory
may clarify this point of view. The transverse components
of the vector potential may be considered as
the gauge-invariant true observables; or they may be
considered as the components of the vector potential
in a particular gauge, namely the radiation gauge.)\footnote{The similarities to gauge theory, and the relation between gauge-invariance and gauge-fixing, has been fleshed out in \citep{Samediff_1b, GomesButterfield_electro}.}
\end{quote}

So let $\phi^{(A)}$ represent a collection of dynamical fields in our models; this could be the metric, or some other set of fields. We fix an isomorphism-class representative of $\phi$ by postulating some constraint $F(\phi^{(A)})\in C^\infty(M)$, such that:
\begin{equation}\label{gf}  \forall \phi^{(A)}(p), \exists ! f_\phi\in Diff(M)\quad |\quad F(f_\phi^*\phi^{(A)}(p))=0.
\end{equation}
then, given any $\phi^{(A)}$, the action of $f_\phi$ will take that model to the \textit{unique}, preferred  representation, $ f_\phi^*\phi^{(A)}$.
Clearly, from uniqueness, for any $d\in \text{Diff}(M)$:
\begin{equation}
    f_{d^*\phi}=d^{-1}f_\phi.
\end{equation}

Now, let 
\begin{equation}
    (\phi^{(A)})_F:=f_\phi^*\phi^{(A)}.
\end{equation}
The previous two equations (and the fact that $(d\circ f)^*=f^*\circ d^*$)  imply that 
\begin{equation}
    (d^*\phi^{(A)})_F=f_\phi^*(d^{-1}\circ d)^*\phi^{(A)}=\phi^{(A)}_F.
\end{equation}
 This shows that   $f_\phi^*\phi^{(A)}$ is invariant under diffeomorphisms: it is what is often called a \emph{relational observable}, obtained by `dressing\rq{} the original fields   $\phi^{(A)}$ with some field-dependent diffeomorphism. Though we found   this dressed observable by employing a gauge-fixing, we need not think of it in those terms: a dressed quantity such as $ f_\phi^*\phi^{(A)}$ is just invariant under diffeomorphisms, full stop.

 Each choice of physical coordinate system, or rather, of physical correlates of spacetime points, can be seen as a realiser of Einstein\rq{}s `point-coincidences\rq{} argument;  each one answers the follow-up question that Einstein never pursued: point coincidences of what?
 
 Moreover, in the cases discussed above, or in the case of Komar variables discussed in Section \ref{sec:Komar}, we can qualitatively individuate spacetime \emph{regions} by their coordinates in a  physical coordinate system.\footnote{And, as we will see in Section \ref{sec:comp}, we could have obtained complete descriptions of the spacetime metric that are relational and invariant under isomorphism and yet do not give criteria for the individuation of spacetime regions that are applicable for all metrics (or all isomorphism classes of the metric). Using such invariants, we would not be able to formulate Noether\rq{}s second theorem, which requires variational symmetries. But the kind of complete, \emph{local}, physical  description of the spacetime metric that we have studied here applies to all (or many) isomorphism classes. \label{ftnt:inv}} Within such descriptions,   location is expressed with respect to some such physical reference, and that location of course varies when the physical reference varies. This variability has recently elicited worries about a new kind of hole argument, sometimes called `a quantum hole argument\rq{} (cf. \cite{ Kabel_et_al2024, AdlamLinnemannRead}). 
 
 For instance, suppose that we have agreed to represent a region of spacetime around the Earth through a set of GPS coordinates. \cite{Rovelli_GPS}  writes down  the emerging relational form of the metric, on a par with \eqref{eq:gbar}, and says (Ibid, p.5): 
 \begin{quote} \lq\lq{}Finally, notice that the observables we have defined are a straightforward generalization of Einstein’s \lq\lq{}point coincidences\rq\rq{}. In a sense, they are precisely Einstein’s point coincidences. Einstein’s \lq\lq{}material points\rq\rq{} are just replaced by photons (light pulses): [a spacetime point] is characterized as the meeting point of four photons\rq\rq{}\end{quote}

 But we could have chosen a second set of GPS coordinates based on a different set of GPS satellites.      Translating between these choices is non-trivial, because the image and domain of the translation use different sets of physical properties to describe locations.   But since each such set of invariant properties and relations is complete---in the sense of uniquely characterizing all the locations in a given region of spacetime---they will be inter-translatable: there is always a 1-1 map that, as described in either set of coordinates, is smooth,  has a smooth inverse, and will bring the descriptions of the fields into coincidence.\footnote{Explicitly, suppose we have two choices of physical correlates for the spacetime points of some region $U\subset M$---two choices of a physical reference frame, $\bar x^\mu$ and $\tilde x^\nu$, as described in e.g. \eqref{eq:Komar_coords} for Komar coordinates, or in \cite{Rovelli_GPS} for GPS coordinates. So, as in Section \ref{sec:active-passive}, we have two diffeomorphisms $\bar \phi, \tilde \phi: U\rightarrow \RR^4$, which we can compose as: 
\be \bar f:= \bar \phi\circ \tilde \phi^{-1}\in \text{Diff}(\RR^4).
\ee
And we have two corresponding forms for the metric, as in \eqref{eq:gbar}: $ g_{\bar \mu \bar \nu},  g_{\tilde \mu \tilde \nu}$. So we get: 
\be g_{\bar \mu \bar \nu}(\bar x)=\frac{\partial f^{\tilde \mu}}{\partial \bar x^\mu}\frac{\partial f^{\tilde \nu}}{\partial \bar x^\nu}g_{\tilde \mu \tilde \nu}(f(\bar x)).
\ee
}

 Can these different choices lead to indeterminism? 
No, it cannot. 
As long as we fix a convention for what is to play the role of a physical reference frame, the evolution of the metric and other fields dynamically coupled to the metric is unique. And it is not unique in the sense that we could have chosen otherwise. For example, the motion of the GPS coordinates is fixed by their initial conditions and equations of motion; we can switch from one to the other, but that will entail a change in the initial conditions (i.e. $\Delta$ in Section \ref{sec:hole}), and the new initial conditions will still uniquely determine the evolution of the metric and other fields. 


The physical coordinate systems are neither abstract  nor arbitrarily redefinable.  By being anchored to physical systems,  they can be unambiguously fixed and, given initial conditions, they will uniquely describe the evolution of the fields, thereby avoiding the indeterminism threatened by the hole argument.


This resolution of the tension between Noether\rq{}s theorem, the Lie derivative and determinism relied on a physical description of the geometry, enabled by the relations between  fields coupled to the spacetime metric. Indeed,  as I will now show, when the descriptions are not relational in this sense, indeterminism creeps back in.


\subsection{Indeterminism strikes back}\label{sec:comp}
Now I will describe two possible obstructions to the argument from the previous Section. The more important obstruction, to be described in Section \ref{sec:rel_var}, arises when the `physical coordinate system\rq{} is physical only in name; for instance, when it is not dynamically coupled to the metric. In that case, we can still use these \lq{}non-material coordinate systems\rq{}---IRF\rq{}s, in the classification of \citep{BamontiRF}---to define a non-trivial Lie derivative. But the argument of the previous Section---that coordinate symmetries articulated in terms of of physical correlates for the spacetime points do not lead to indeterminism---fails to hold here. 

The second obstruction, described in Section \ref{sec:obs_local}, arises from having relational descriptions of the physical states that do not allow us to individuate regions for different physical states. That is, conceptually, our unique description of the isomorphism-related models via a set of qualitative properties or relations is underwritten by anti-haecceitism, because the properties and relations encoded in the dressed field $(\phi^{(A)})_F$ are insensitive only  to the haecceitistic differences between  models. But, as described briefly in footnote \ref{ftnt:inv}, invariance is not enough to obtain vector fields that are qualitative, or physical, and persist across a neighborhood of isomorphism classes in the space of models. In other words, in order to apply the variational methods of Noether\rq{}s second theorem, in Section \ref{sec:Noether}, we need physical vector fields definable on open neighborhoods of $g_{ab}$ in the space of Lorentzian metrics.\footnote{Or rather, the vector fields need to be defined on  open neighborhoods of orbits around the isomorphism-orbit of $g_{ab}$. }

To articulate both of these obstructions,  we first suppose that the $\phi^{(A)}$ of the previous Section describes  four scalar fields, $\phi^{(A)}, A=1, \cdots, 4$, with which we describe a (local) diffeomorphism $ U\rightarrow \mathbb{R}^4$ with $U\subset M$, and a metric  $g_{ab}$. 
Let us assume that in fact $M$ is diffeomorphic to $\mathbb{R}^4$, so that we can choose $U=M$. Though this assumption cannot be generically upheld, it serves to illustrate my claims.\footnote{
It is important to note a limitation of such an assumption: any two values for the scalar fields are related by a diffeomorphism of $M$, since $(\phi^{(A)})^{-1}\circ \phi'^{(A)}\in Diff(M) $. }

Given \textit{any} doublet: $(g_{ab}, \phi)$, the composition $g_{ab}\circ (\phi^{(A)})^{-1}$, defined by the components $g_{AB}$ in this physical coordinate system, is  invariant under diffeomorphisms. That is because, given $p=(\phi^{(A)})^{-1}(x) \in M$, 
\be (f^*\phi^{(A)})^{-1}(x)=f^{-1}(p),\quad  \text{and}\quad g_{ab}(p)=f^*g_{ab}(f^{-1}(p)),\ee
so $g_{ab}\circ (\phi^{(A)})^{-1}=f^*g_{ab}\circ (f^*\phi^{(A)})^{-1}$.
As expected, both fields change together, so that the relations between them do not change.

\subsubsection{The first obstruction: indeterminism}\label{sec:rel_var}
To see the first obstruction,  we first recall the good case, studied in Section \ref{sec:Komar}:  there, $g_{ab}$ and $\phi$ or $\mathcal{R}^\mu_g$ are fully dynamically coupled, so that when e.g. $(g_{ab},\phi^{(A)})$ is a dynamically possible model,  $(f^*g_{ab},\phi^{(A)})$ \emph{is not}. For instance, suppose that each of the $\phi^{(A)}$ satisfies a wave equation 
\be \square\phi^{(A)}=0,
\ee 
where $\square$ is the D\rq{}Alembertian according to $g_{ab}$ and the four fields satisfy linearly independent boundary conditions. 
This case is similar to that discussed in the previous Section, in which the physical coordinate system was constructed from the metric itself.

 Moving on to the second, problematic case,  using a different field to make up the reference frame, unlike the case with the $\mathcal{R}_g$, we can suppose $g_{ab}$ and $\phi^{(A)}$ \emph{aren't} dynamically coupled. 
 We can, that is, suppose that the two kinds of fields are completely independent.  In this case, the values of $\phi^{(A)}$---a `farie' or `ghostly' field as far as the metric $g_{ab}$ is concerned---are still compatible with every value of  $g_{ab}$.  
In other words, if $g_{ab}$ and $\phi$ are \textit{not dynamically coupled}, any $\phi^{(A)}$ is still compatible with any of  the isomorphic metrics, i.e. $\mathcal{M}_1:=\langle M, g_{ab}, \phi^{(A)}\rangle$ and $\mathcal{M}_2:=\langle M, f^*g_{ab}, \phi^{(A)}\rangle$ are legitimate models for the dynamics, for for all  $f \in \text{Diff}(M)$.

 In this case, while it is true that $g_{ab}\circ (\phi^{(A)})^{-1}$ is diffeomorphism invariant, so is $f^*g_{ab}\circ (\phi^{(A)})^{-1}$, but we have no reason to start with one $g_{ab}$ rather than with an isomorphic metric.
 If we use a gauge-fixed  $\phi^{(A)}$ (or some physical properties of $\phi^{(A)}$) to define physical curves on spacetime, we can once again define the Lie derivative, as in the previous Section. But the hole argument still threatens indeterminism, since given initial data for the metric, the equations of motion determine only the entire isomorphism class.\footnote{ 
This point is often overlooked in the literature about \emph{partial and complete observables}, cf. \citep{Rovelli_partial}. The main idea behind these terms is to relate different sets of gauge-dependent fields (partial observables) in a gauge-invariant manner, thus constructing a `complete observable\rq{} by composition.
This construction implements the general idea that the physical content of GR lies in the relations between dynamic quantities represented by partial observables. In other words, the general idea is that we observe relational evolution between fields and not evolution with respect to some background unobservable structure.}

Let me make the point more directly. Suppose you worry that GR is indeterministic because you have hole-diffeomorphic-related metrics. Then you say: well, I'll get rid of redundancy by finding diffeomorphism-invariant observables. That is fine, and it would usually work.  

But if you pursue this strategy by invoking  fields that are uncoupled to the metric as reference frames, diffeomorphism-invariance comes apart from the notion of geometric significance that underwrites  the usual anti-haecceitist response to the hole argument. 

That is,  a reshuffling of points of the manifold will equally affect all of the fields: under its action  $\mathcal{M}_1:=\langle M, g_{ab}, \phi^{(A)}\rangle$ is only on the same isomorphism class as $\mathcal{M}\rq{}_1:=\langle M, f^*g_{ab},f^*\phi^{(A)}\rangle$, but it is in a different isomorphism class as $\mathcal{M}_2:=\langle M, f^*g_{ab},\phi^{(A)}\rangle$: the doublet of fields `relationally differ\rq{} in the two alternatives. No reshuffling of points can bring one model  to the other. The accompanying interpretation would be that  $\mathcal{M}_1$ and $\mathcal{M}_2$ represent physically distinct states for its target spacetime regions, because they involve different physical relations. And yet,  supposing that the equations of motion of $\phi$ are also invariant under diffeomorphisms---say, they satisfy $\square_h \phi^{(A)}=0$, where $\square_h$ is the D\rq{}Alembertian according to an auxiliary fixed, background metric, $h_{ab}$--- the two models would be related by a dynamical symmetry: here we can act with independent diffeomorphisms on the metric and on the matter fields, while still preserving the equations of motion. In other words, here dynamical symmetries---which allow a separate action of the diffeomorphisms on both $\phi$ and $g_{ab}$---have come apart from isomorphisms in quite a radical way: there are physical relations that are not preserved by the dynamical symmetries. 
  By dynamically coupling $g_{ab}$ to $\phi$, as we did in the previous Section,  we  disallow one of the two  isomorphism classes---either $\mathcal{M}_1$ or $\mathcal{M}_2$--- as being dynamically allowed, and thereby we recover determinism.

 In sum,  in order to unambiguously respond to the threat of indeterminism as presented by the hole argument, we need to construct invariant quantities using dynamically coupled fields. If they are uncoupled, we can still represent both hole-diffeomorphic-related metrics in a `physical\rq{} reference frame; and indeterminism still looms. But as we will now discuss, having dynamically coupled fields does not suffice to obtain the necessary ingredients for our physical understanding of Noether\rq{}s second theorem.

 \subsubsection{Not just relational: local.}\label{sec:obs_local}
 Now onto the second possible obstruction to our resolution of the hole argument compatible with Noether\rq{}s second theorem. An application of Noether\rq{}s theorem requires us to physically individuate directions in $M$ across neighborhoods  \emph{in the space of models} of general relativity. In other words, since Noether\rq{}s second theorem, as described in Section \ref{sec:Noether} (see equation \eqref{eq:2_var}), is variational, we need to have vector fields $X$ that are constructed by physical properties of metrics (and possibly other fields), across open sets of non-isomorphic metrics. 

Agreed: as in the introduction to this Section, given \textit{any} doublet, $(g_{ab}, \phi)$, the composition $g_{AB}(\phi):=g_{ab}\circ (\phi^{(A)})^{-1}$  is  invariant under diffeomorphisms. While it is true that a range of values of $g_{AB}(\phi)$ would single out a region of $M$, this would only single out the region in the isomorphism class of $\mathcal{M}_1:=\langle M, g_{ab}, \phi^{(A)}\rangle$. A neighboring, different isomorphism class, of $\mathcal{M}_2$, would not have those values of $g_{AB}(\phi)$ and so could not single out any region by that criterion.\footnote{Incidentally, this is the same problem faced by \citet{Maudlin_essence}\rq{}s \emph{metric essentialist} response to the hole argument. There, each spacetime point has its metric properties essentially; so that isomorphic-related possibilities are ruled out. But essentialism forbids other isomorphism classes, in which the same point of $M$ would have other, non-isomorphic, distribution of metric properties and relations. }

Note also that each  $\phi^{(A)}$ in an isomorphism class may correspond to a different vector field on $M$, so, naively, we cannot define the vector field by the entire isomorphism class of $\phi^{(A)}$. The solution of this problem, explored in \ref{sec:Komar}, was to pick out a certain kind of invariant, formed by relational properties of the field $\phi^{(A)}$, by methods similar to gauge-fixing.

\section{Summary}

The threat of indeterminism posed by the hole argument is satisfactorily thwarted by an anti-haecceitist understanding of fields on a spacetime manifold, as described in Section \ref{sec:drag}. In this understanding, the difference between  isomorphic related models concerns only `which point fills which qualitative properties and relations\rq{}, and these are haecceitistic differences, rejected by the anti-haecceitist. 

One way to try to `wear anti-haecceitism\rq{} on our sleeves, so to speak, is to endorse `the drag-along\rq{} understanding of isomorphisms, the basis for the drag-along response to the hole argument. According to such an understanding, the only mathematically legitimate way to compare isomorphic models is via the isomorphism that relates them. 

But this `drag-along\rq{} understanding of isomorphism-related models lies in tension with the  standards of mathematical practice in general relativity, as found in all textbooks. For it is very common to compare isometric models by  diffeomorphisms other than the one that gives rise to the isometry (see \citep{GomesButterfield_hole1}). 

These other standards of comparisons are used, for instance,   in the proof of  Noether\rq{}s second theorem. Given two conditions on the coupling of fields to the geometry (essentially, minimal coupling), the theorem concludes that the energy-momentum tensor is locally (covariantly) conserved. With only the drag-along notion of comparing isomorphic spacetimes we lose this line of argument entirely. 

Here I resolved this tension, with a three-step argument. First, I constructed diffeomorphisms that necessarily map between spacetime points with distinct qualitative profiles, from which I could unproblematically define non-trivial Lie derivatives. Second, I showed that these diffeomorphisms could be understood as changes between physical coordinate systems, and so constituted symmetries of the Einstein equations of motion in a coordinate basis. Nonetheless, as showed in the third step of the argument,  there is no threat of indeterminism: these are physical coordinate systems, and each choice uniquely defines a relational evolution of the metric.

Indeed, in order to avoid indeterminism, the physical content of the coordinate systems is essential: the fields out of which we build the coordinates must be physically coupled to the metric. Had we posited an abstract coordinate system made out of  fields that did not dynamically couple to the metric,  the evolution of the metric would have remained (perniciously) indeterministic, even if defined in a relational diffeomorphism-invariant manner. And had we required a coordinate system that had no physical significance on its own, we might still be able to write down diffeomorphism-invariant components of the metric, but that would not suffice to construct the physical directions in spacetime that are necessary for an application of Noether\rq{}s second theorem.

\subsection*{Acknowledgements} I would like to thank Jeremy Butterfield, Oliver Pooley, and Nicola Bamonti, for many discussions and suggestions. 

\bibliographystyle{apacite} 
\bibliography{references3}

\begin{thebibliography}{}

\bibitem [\protect \citeauthoryear {%
Adlam%
, Linnemann%
\BCBL {}\ \BBA {} Read%
}{%
Adlam%
\ \protect \BOthers {.}}{%
{\protect \APACyear {2022}}%
}]{%
AdlamLinnemannRead}
\APACinsertmetastar {%
AdlamLinnemannRead}%
\begin{APACrefauthors}%
Adlam, E.%
, Linnemann, N.%
\BCBL {}\ \BBA {} Read, J.%
\end{APACrefauthors}%
\unskip\
\newblock
\APACrefYearMonthDay{2022}{}{}.
\newblock
{\BBOQ}\APACrefatitle {{Constructive Axiomatics in Spacetime Physics Part II:
  Constructive Axiomatics in Context}} {{Constructive Axiomatics in Spacetime
  Physics Part II: Constructive Axiomatics in Context}}.{\BBCQ}
\newblock
\APACjournalVolNumPages{In preparation}{}{}{}.
\PrintBackRefs{\CurrentBib}

\bibitem [\protect \citeauthoryear {%
Alvarez-Gaumé%
\ \BBA {} Witten%
}{%
Alvarez-Gaumé%
\ \BBA {} Witten%
}{%
{\protect \APACyear {1984}}%
}]{%
AlvarezWitten}
\APACinsertmetastar {%
AlvarezWitten}%
\begin{APACrefauthors}%
Alvarez-Gaumé, L.%
\BCBT {}\ \BBA {} Witten, E.%
\end{APACrefauthors}%
\unskip\
\newblock
\APACrefYearMonthDay{1984}{}{}.
\newblock
{\BBOQ}\APACrefatitle {Gravitational anomalies} {Gravitational
  anomalies}.{\BBCQ}
\newblock
\APACjournalVolNumPages{Nuclear Physics B}{234}{2}{269-330}.
\newblock
\begin{APACrefURL}
  \url{https://www.sciencedirect.com/science/article/pii/055032138490066X}
  \end{APACrefURL}
\newblock
\begin{APACrefDOI} \doi{https://doi.org/10.1016/0550-3213(84)90066-X}
  \end{APACrefDOI}
\PrintBackRefs{\CurrentBib}

\bibitem [\protect \citeauthoryear {%
Bamonti%
}{%
Bamonti%
}{%
{\protect \APACyear {2023}}%
}]{%
BamontiRF}
\APACinsertmetastar {%
BamontiRF}%
\begin{APACrefauthors}%
Bamonti, N.%
\end{APACrefauthors}%
\unskip\
\newblock
\APACrefYearMonthDay{2023}{}{}.
\newblock
\APACrefbtitle {What is a reference frame in General Relativity?} {What is a
  reference frame in general relativity?}
\PrintBackRefs{\CurrentBib}

\bibitem [\protect \citeauthoryear {%
Bradley%
\ \BBA {} Weatherall%
}{%
Bradley%
\ \BBA {} Weatherall%
}{%
{\protect \APACyear {2022}}%
}]{%
BradleyWeatherall_hole}
\APACinsertmetastar {%
BradleyWeatherall_hole}%
\begin{APACrefauthors}%
Bradley, C.%
\BCBT {}\ \BBA {} Weatherall, J\BPBI O.%
\end{APACrefauthors}%
\unskip\
\newblock
\APACrefYearMonthDay{2022}{}{}.
\newblock
{\BBOQ}\APACrefatitle {Mathematical Responses to the Hole Argument: Then and
  Now} {Mathematical responses to the hole argument: Then and now}.{\BBCQ}
\newblock
\APACjournalVolNumPages{Philosophy of Science}{89}{5}{1223–1232}.
\newblock
\begin{APACrefDOI} \doi{10.1017/psa.2022.58} \end{APACrefDOI}
\PrintBackRefs{\CurrentBib}

\bibitem [\protect \citeauthoryear {%
Cudek%
}{%
Cudek%
}{%
{\protect \APACyear {2024}}%
}]{%
Cudek_hole}
\APACinsertmetastar {%
Cudek_hole}%
\begin{APACrefauthors}%
Cudek, F.%
\end{APACrefauthors}%
\unskip\
\newblock
\APACrefYearMonthDay{2024}{}{}.
\newblock
{\BBOQ}\APACrefatitle {Counterparts, Determinism, and the Hole Argument}
  {Counterparts, determinism, and the hole argument}.{\BBCQ}
\newblock
\APACjournalVolNumPages{British Journal for the Philosophy of Science}{}{}{}.
\newblock
\begin{APACrefDOI} \doi{10.1086/729767} \end{APACrefDOI}
\PrintBackRefs{\CurrentBib}

\bibitem [\protect \citeauthoryear {%
Curiel%
}{%
Curiel%
}{%
{\protect \APACyear {2018}}%
}]{%
Curiel2018}
\APACinsertmetastar {%
Curiel2018}%
\begin{APACrefauthors}%
Curiel, E.%
\end{APACrefauthors}%
\unskip\
\newblock
\APACrefYearMonthDay{2018}{}{}.
\newblock
{\BBOQ}\APACrefatitle {{On the Existence of Spacetime Structure}} {{On the
  Existence of Spacetime Structure}}.{\BBCQ}
\newblock
\APACjournalVolNumPages{The British Journal for the Philosophy of
  Science}{69}{2}{447-483}.
\newblock
\begin{APACrefURL} \url{https://doi.org/10.1093/bjps/axw014} \end{APACrefURL}
\newblock
\begin{APACrefDOI} \doi{10.1093/bjps/axw014} \end{APACrefDOI}
\PrintBackRefs{\CurrentBib}

\bibitem [\protect \citeauthoryear {%
Earman%
\ \BBA {} Norton%
}{%
Earman%
\ \BBA {} Norton%
}{%
{\protect \APACyear {1987}}%
}]{%
EarmanNorton1987}
\APACinsertmetastar {%
EarmanNorton1987}%
\begin{APACrefauthors}%
Earman, J.%
\BCBT {}\ \BBA {} Norton, J.%
\end{APACrefauthors}%
\unskip\
\newblock
\APACrefYearMonthDay{1987}{12}{}.
\newblock
{\BBOQ}\APACrefatitle {{What Price Spacetime Substantivalism? The Hole Story}}
  {{What Price Spacetime Substantivalism? The Hole Story}}.{\BBCQ}
\newblock
\APACjournalVolNumPages{The British Journal for the Philosophy of
  Science}{38}{4}{515-525}.
\newblock
\begin{APACrefURL} \url{https://doi.org/10.1093/bjps/38.4.515} \end{APACrefURL}
\newblock
\begin{APACrefDOI} \doi{10.1093/bjps/38.4.515} \end{APACrefDOI}
\PrintBackRefs{\CurrentBib}

\bibitem [\protect \citeauthoryear {%
Einstein%
}{%
Einstein%
}{%
{\protect \APACyear {1916}}%
}]{%
Einstein_points}
\APACinsertmetastar {%
Einstein_points}%
\begin{APACrefauthors}%
Einstein, A.%
\end{APACrefauthors}%
\unskip\
\newblock
\APACrefYearMonthDay{1916}{}{}.
\newblock
{\BBOQ}\APACrefatitle {{ Grundlage der allgemeinen Relativit\"atstheorie}} {{
  Grundlage der allgemeinen Relativit\"atstheorie}}.{\BBCQ}
\newblock
\APACjournalVolNumPages{Annalen der Physik, 49. pp 769-822}{}{}{}.
\PrintBackRefs{\CurrentBib}

\bibitem [\protect \citeauthoryear {%
Field%
}{%
Field%
}{%
{\protect \APACyear {1984}}%
}]{%
Field_soph}
\APACinsertmetastar {%
Field_soph}%
\begin{APACrefauthors}%
Field, H.%
\end{APACrefauthors}%
\unskip\
\newblock
\APACrefYearMonthDay{1984}{}{}.
\newblock
{\BBOQ}\APACrefatitle {Can We Dispense with Space-Time?} {Can we dispense with
  space-time?}{\BBCQ}
\newblock
\APACjournalVolNumPages{PSA: Proceedings of the Biennial Meeting of the
  Philosophy of Science Association}{1984}{}{33--90}.
\newblock
\begin{APACrefURL} [{2022-08-30}]\url{http://www.jstor.org/stable/192496}
  \end{APACrefURL}
\PrintBackRefs{\CurrentBib}

\bibitem [\protect \citeauthoryear {%
Fischer%
\ \BBA {} Marsden%
}{%
Fischer%
\ \BBA {} Marsden%
}{%
{\protect \APACyear {1979}}%
}]{%
fischermarsden}
\APACinsertmetastar {%
fischermarsden}%
\begin{APACrefauthors}%
Fischer, A\BPBI E.%
\BCBT {}\ \BBA {} Marsden, J\BPBI E.%
\end{APACrefauthors}%
\unskip\
\newblock
\APACrefYearMonthDay{1979}{}{}.
\newblock
{\BBOQ}\APACrefatitle {The initial value problem and the dynamical formulation
  of general relativity.} {The initial value problem and the dynamical
  formulation of general relativity.}{\BBCQ}
\newblock
\BIn{} \APACrefbtitle {General relativity : an Einstein centenary survey.
  Cambridge University Press , New York, pp. 138-211.} {General relativity : an
  einstein centenary survey. cambridge university press , new york, pp.
  138-211.}
\PrintBackRefs{\CurrentBib}

\bibitem [\protect \citeauthoryear {%
Fletcher%
}{%
Fletcher%
}{%
{\protect \APACyear {2020}}%
}]{%
Fletcher_hole}
\APACinsertmetastar {%
Fletcher_hole}%
\begin{APACrefauthors}%
Fletcher, S\BPBI C.%
\end{APACrefauthors}%
\unskip\
\newblock
\APACrefYearMonthDay{2020}{}{}.
\newblock
{\BBOQ}\APACrefatitle {{On Representational Capacities, with an Application to
  General Relativity}} {{On Representational Capacities, with an Application to
  General Relativity}}.{\BBCQ}
\newblock
\APACjournalVolNumPages{Foundations of Physics}{50}{4}{228--249}.
\newblock
\begin{APACrefDOI} \doi{10.1007/s10701-018-0208-6} \end{APACrefDOI}
\PrintBackRefs{\CurrentBib}

\bibitem [\protect \citeauthoryear {%
Giovanelli%
}{%
Giovanelli%
}{%
{\protect \APACyear {2021}}%
}]{%
Giovanelli2021}
\APACinsertmetastar {%
Giovanelli2021}%
\begin{APACrefauthors}%
Giovanelli, M.%
\end{APACrefauthors}%
\unskip\
\newblock
\APACrefYearMonthDay{2021}{may}{}.
\newblock
{\BBOQ}\APACrefatitle {{Nothing but coincidences: the point-coincidence and
  Einstein's struggle with the meaning of coordinates in physics}} {{Nothing
  but coincidences: the point-coincidence and Einstein's struggle with the
  meaning of coordinates in physics}}.{\BBCQ}
\newblock
\APACjournalVolNumPages{European Journal for Philosophy of Science}{11}{2}{}.
\newblock
\begin{APACrefDOI} \doi{10.1007/s13194-020-00332-7} \end{APACrefDOI}
\PrintBackRefs{\CurrentBib}

\bibitem [\protect \citeauthoryear {%
Gomes%
}{%
Gomes%
}{%
{\protect \APACyear {2021}}%
}]{%
Samediff_1a}
\APACinsertmetastar {%
Samediff_1a}%
\begin{APACrefauthors}%
Gomes, H.%
\end{APACrefauthors}%
\unskip\
\newblock
\APACrefYearMonthDay{2021}{}{}.
\newblock
{\BBOQ}\APACrefatitle {{Same-diff? Conceptual similarities between gauge
  transformations and diffeomorphisms. Part II: Challenges to sophistication}}
  {{Same-diff? Conceptual similarities between gauge transformations and
  diffeomorphisms. Part II: Challenges to sophistication}}.{\BBCQ}
\newblock
\APACjournalVolNumPages{Arxiv: 2110.07204. Submitted.}{}{}{}.
\PrintBackRefs{\CurrentBib}

\bibitem [\protect \citeauthoryear {%
Gomes%
}{%
Gomes%
}{%
{\protect \APACyear {2022}}%
}]{%
Samediff_1b}
\APACinsertmetastar {%
Samediff_1b}%
\begin{APACrefauthors}%
Gomes, H.%
\end{APACrefauthors}%
\unskip\
\newblock
\APACrefYearMonthDay{2022}{}{}.
\newblock
{\BBOQ}\APACrefatitle {{Same-diff? Conceptual similarities between gauge
  transformations and diffeomorphisms. Part III: Representational conventions
  and relationism}} {{Same-diff? Conceptual similarities between gauge
  transformations and diffeomorphisms. Part III: Representational conventions
  and relationism}}.{\BBCQ}
\newblock
\APACjournalVolNumPages{Unpublished}{}{}{}.
\PrintBackRefs{\CurrentBib}

\bibitem [\protect \citeauthoryear {%
Gomes%
\ \BBA {} Butterfield%
}{%
Gomes%
\ \BBA {} Butterfield%
}{%
{\protect \APACyear {2022}}%
}]{%
GomesButterfield_electro}
\APACinsertmetastar {%
GomesButterfield_electro}%
\begin{APACrefauthors}%
Gomes, H.%
\BCBT {}\ \BBA {} Butterfield, J.%
\end{APACrefauthors}%
\unskip\
\newblock
\APACrefYearMonthDay{2022}{aug}{}.
\newblock
{\BBOQ}\APACrefatitle {{How to Choose a Gauge? The Case of Hamiltonian
  Electromagnetism}} {{How to Choose a Gauge? The Case of Hamiltonian
  Electromagnetism}}.{\BBCQ}
\newblock
\APACjournalVolNumPages{Erkenntnis}{}{}{}.
\newblock
\begin{APACrefDOI} \doi{10.1007/s10670-022-00597-9} \end{APACrefDOI}
\PrintBackRefs{\CurrentBib}

\bibitem [\protect \citeauthoryear {%
Gomes%
\ \BBA {} Butterfield%
}{%
Gomes%
\ \BBA {} Butterfield%
}{%
{\protect \APACyear {2023}}%
{\protect \APACexlab {{\protect \BCnt {1}}}}}]{%
GomesButterfield_hole2}
\APACinsertmetastar {%
GomesButterfield_hole2}%
\begin{APACrefauthors}%
Gomes, H.%
\BCBT {}\ \BBA {} Butterfield, J.%
\end{APACrefauthors}%
\unskip\
\newblock
\APACrefYearMonthDay{2023{\protect \BCnt {1}}}{jun}{}.
\newblock
{\BBOQ}\APACrefatitle {{The Hole Argument and Beyond: Part II: Treating
  Non-isomorphic Spacetimes}} {{The Hole Argument and Beyond: Part II: Treating
  Non-isomorphic Spacetimes}}.{\BBCQ}
\newblock
\APACjournalVolNumPages{Journal of Physics: Conference
  Series}{2533}{1}{012003}.
\newblock
\begin{APACrefURL} \url{https://dx.doi.org/10.1088/1742-6596/2533/1/012003}
  \end{APACrefURL}
\newblock
\begin{APACrefDOI} \doi{10.1088/1742-6596/2533/1/012003} \end{APACrefDOI}
\PrintBackRefs{\CurrentBib}

\bibitem [\protect \citeauthoryear {%
Gomes%
\ \BBA {} Butterfield%
}{%
Gomes%
\ \BBA {} Butterfield%
}{%
{\protect \APACyear {2023}}%
{\protect \APACexlab {{\protect \BCnt {2}}}}}]{%
GomesButterfield_hole1}
\APACinsertmetastar {%
GomesButterfield_hole1}%
\begin{APACrefauthors}%
Gomes, H.%
\BCBT {}\ \BBA {} Butterfield, J.%
\end{APACrefauthors}%
\unskip\
\newblock
\APACrefYearMonthDay{2023{\protect \BCnt {2}}}{jun}{}.
\newblock
{\BBOQ}\APACrefatitle {{The Hole Argument and Beyond: Part I: The Story so
  Far}} {{The Hole Argument and Beyond: Part I: The Story so Far}}.{\BBCQ}
\newblock
\APACjournalVolNumPages{Journal of Physics: Conference
  Series}{2533}{1}{012002}.
\newblock
\begin{APACrefURL} \url{https://dx.doi.org/10.1088/1742-6596/2533/1/012002}
  \end{APACrefURL}
\newblock
\begin{APACrefDOI} \doi{10.1088/1742-6596/2533/1/012002} \end{APACrefDOI}
\PrintBackRefs{\CurrentBib}

\bibitem [\protect \citeauthoryear {%
Gomes%
, Hopfmüller%
\BCBL {}\ \BBA {} Riello%
}{%
Gomes%
\ \protect \BOthers {.}}{%
{\protect \APACyear {2019}}%
}]{%
GomesHopfRiello}
\APACinsertmetastar {%
GomesHopfRiello}%
\begin{APACrefauthors}%
Gomes, H.%
, Hopfmüller, F.%
\BCBL {}\ \BBA {} Riello, A.%
\end{APACrefauthors}%
\unskip\
\newblock
\APACrefYearMonthDay{2019}{}{}.
\newblock
{\BBOQ}\APACrefatitle {A unified geometric framework for boundary charges and
  dressings: Non-Abelian theory and matter} {A unified geometric framework for
  boundary charges and dressings: Non-abelian theory and matter}.{\BBCQ}
\newblock
\APACjournalVolNumPages{Nuclear Physics B}{941}{}{249 - 315}.
\newblock
\begin{APACrefURL}
  \url{http://www.sciencedirect.com/science/article/pii/S0550321319300483}
  \end{APACrefURL}
\newblock
\begin{APACrefDOI} \doi{https://doi.org/10.1016/j.nuclphysb.2019.02.020}
  \end{APACrefDOI}
\PrintBackRefs{\CurrentBib}

\bibitem [\protect \citeauthoryear {%
Gomes%
\ \BBA {} Riello%
}{%
Gomes%
\ \BBA {} Riello%
}{%
{\protect \APACyear {2021}}%
}]{%
GomesRiello_new}
\APACinsertmetastar {%
GomesRiello_new}%
\begin{APACrefauthors}%
Gomes, H.%
\BCBT {}\ \BBA {} Riello, A.%
\end{APACrefauthors}%
\unskip\
\newblock
\APACrefYearMonthDay{2021}{}{}.
\newblock
{\BBOQ}\APACrefatitle {{The quasilocal degrees of freedom of Yang-Mills
  theory}} {{The quasilocal degrees of freedom of Yang-Mills theory}}.{\BBCQ}
\newblock
\APACjournalVolNumPages{SciPost Phys.}{10}{}{130}.
\newblock
\begin{APACrefURL} \url{https://scipost.org/10.21468/SciPostPhys.10.6.130}
  \end{APACrefURL}
\newblock
\begin{APACrefDOI} \doi{10.21468/SciPostPhys.10.6.130} \end{APACrefDOI}
\PrintBackRefs{\CurrentBib}

\bibitem [\protect \citeauthoryear {%
Hoefer%
}{%
Hoefer%
}{%
{\protect \APACyear {1996}}%
}]{%
Hoefer_hole}
\APACinsertmetastar {%
Hoefer_hole}%
\begin{APACrefauthors}%
Hoefer, C.%
\end{APACrefauthors}%
\unskip\
\newblock
\APACrefYearMonthDay{1996}{}{}.
\newblock
{\BBOQ}\APACrefatitle {{The Metaphysics of Space-Time Substantivalism}} {{The
  Metaphysics of Space-Time Substantivalism}}.{\BBCQ}
\newblock
\APACjournalVolNumPages{The Journal of Philosophy}{93}{1}{5--27}.
\newblock
\begin{APACrefURL} \url{http://www.jstor.org/stable/2941016} \end{APACrefURL}
\PrintBackRefs{\CurrentBib}

\bibitem [\protect \citeauthoryear {%
Iftime%
\ \BBA {} Stachel%
}{%
Iftime%
\ \BBA {} Stachel%
}{%
{\protect \APACyear {2006}}%
}]{%
Stachel_Iftime_short}
\APACinsertmetastar {%
Stachel_Iftime_short}%
\begin{APACrefauthors}%
Iftime, M.%
\BCBT {}\ \BBA {} Stachel, J.%
\end{APACrefauthors}%
\unskip\
\newblock
\APACrefYearMonthDay{2006}{}{}.
\newblock
{\BBOQ}\APACrefatitle {The hole argument for covariant theories} {The hole
  argument for covariant theories}.{\BBCQ}
\newblock
\APACjournalVolNumPages{General Relativity and Gravitation {\bf{38}},
  1241–1252}{}{}{}.
\newblock
\begin{APACrefDOI} \doi{10.1007/s10714-006-0303-4} \end{APACrefDOI}
\PrintBackRefs{\CurrentBib}

\bibitem [\protect \citeauthoryear {%
Kabel%
\ \protect \BOthers {.}}{%
Kabel%
\ \protect \BOthers {.}}{%
{\protect \APACyear {2024}}%
}]{%
Kabel_et_al2024}
\APACinsertmetastar {%
Kabel_et_al2024}%
\begin{APACrefauthors}%
Kabel, V.%
, de~la Hamette, A\BHBI C.%
, Apadula, L.%
, Cepollaro, C.%
, Gomes, H.%
, Butterfield, J.%
\BCBL {}\ \BBA {} Časlav Brukner.%
\end{APACrefauthors}%
\unskip\
\newblock
\APACrefYearMonthDay{2024}{}{}.
\newblock
\APACrefbtitle {Identification is Pointless: Quantum Reference Frames,
  Localisation of Events, and the Quantum Hole Argument.} {Identification is
  pointless: Quantum reference frames, localisation of events, and the quantum
  hole argument.}
\PrintBackRefs{\CurrentBib}

\bibitem [\protect \citeauthoryear {%
Kment%
}{%
Kment%
}{%
{\protect \APACyear {2012}}%
}]{%
Kment}
\APACinsertmetastar {%
Kment}%
\begin{APACrefauthors}%
Kment, B.%
\end{APACrefauthors}%
\unskip\
\newblock
\APACrefYearMonthDay{2012}{}{}.
\newblock
{\BBOQ}\APACrefatitle {{Haecceitism, Chance, and Counterfactuals}}
  {{Haecceitism, Chance, and Counterfactuals}}.{\BBCQ}
\newblock
\APACjournalVolNumPages{Philosophical Review}{121}{4}{573--609}.
\newblock
\begin{APACrefDOI} \doi{10.1215/00318108-1630930} \end{APACrefDOI}
\PrintBackRefs{\CurrentBib}

\bibitem [\protect \citeauthoryear {%
Komar%
}{%
Komar%
}{%
{\protect \APACyear {1958}}%
}]{%
Komar_inv}
\APACinsertmetastar {%
Komar_inv}%
\begin{APACrefauthors}%
Komar, A.%
\end{APACrefauthors}%
\unskip\
\newblock
\APACrefYearMonthDay{1958}{Aug}{}.
\newblock
{\BBOQ}\APACrefatitle {{Construction of a Complete Set of Independent
  Observables in the General Theory of Relativity}} {{Construction of a
  Complete Set of Independent Observables in the General Theory of
  Relativity}}.{\BBCQ}
\newblock
\APACjournalVolNumPages{Physical Review}{111}{}{1182--1187}.
\newblock
\begin{APACrefURL} \url{https://link.aps.org/doi/10.1103/PhysRev.111.1182}
  \end{APACrefURL}
\newblock
\begin{APACrefDOI} \doi{10.1103/PhysRev.111.1182} \end{APACrefDOI}
\PrintBackRefs{\CurrentBib}

\bibitem [\protect \citeauthoryear {%
Kretschmann%
}{%
Kretschmann%
}{%
{\protect \APACyear {1918}}%
}]{%
Kretschmann_inv}
\APACinsertmetastar {%
Kretschmann_inv}%
\begin{APACrefauthors}%
Kretschmann, E.%
\end{APACrefauthors}%
\unskip\
\newblock
\APACrefYearMonthDay{1918}{}{}.
\newblock
{\BBOQ}\APACrefatitle {{Über den physikalischen Sinn der
  Relativitätspostulate, A. Einsteins neue und seine ursprüngliche
  Relativitätstheorie}} {{Über den physikalischen Sinn der
  Relativitätspostulate, A. Einsteins neue und seine ursprüngliche
  Relativitätstheorie}}.{\BBCQ}
\newblock
\APACjournalVolNumPages{Annalen der Physik}{358}{16}{575-614}.
\newblock
\begin{APACrefURL}
  \url{https://onlinelibrary.wiley.com/doi/abs/10.1002/andp.19183581602}
  \end{APACrefURL}
\newblock
\begin{APACrefDOI} \doi{https://doi.org/10.1002/andp.19183581602}
  \end{APACrefDOI}
\PrintBackRefs{\CurrentBib}

\bibitem [\protect \citeauthoryear {%
Lee%
\ \BBA {} Wald%
}{%
Lee%
\ \BBA {} Wald%
}{%
{\protect \APACyear {1990}}%
}]{%
Lee:1990nz}
\APACinsertmetastar {%
Lee:1990nz}%
\begin{APACrefauthors}%
Lee, J.%
\BCBT {}\ \BBA {} Wald, R\BPBI M.%
\end{APACrefauthors}%
\unskip\
\newblock
\APACrefYearMonthDay{1990}{}{}.
\newblock
{\BBOQ}\APACrefatitle {{Local symmetries and constraints}} {{Local symmetries
  and constraints}}.{\BBCQ}
\newblock
\APACjournalVolNumPages{J. Math. Phys.}{31}{}{725-743}.
\newblock
\begin{APACrefDOI} \doi{10.1063/1.528801} \end{APACrefDOI}
\PrintBackRefs{\CurrentBib}

\bibitem [\protect \citeauthoryear {%
Lutz%
}{%
Lutz%
}{%
{\protect \APACyear {2017}}%
}]{%
Lutz_syntax}
\APACinsertmetastar {%
Lutz_syntax}%
\begin{APACrefauthors}%
Lutz, S.%
\end{APACrefauthors}%
\unskip\
\newblock
\APACrefYearMonthDay{2017}{}{}.
\newblock
{\BBOQ}\APACrefatitle {{What Was the Syntax-Semantics Debate in the Philosophy
  of Science About?}} {{What Was the Syntax-Semantics Debate in the Philosophy
  of Science About?}}{\BBCQ}
\newblock
\APACjournalVolNumPages{Philosophy and Phenomenological
  Research}{95}{2}{319-352}.
\newblock
\begin{APACrefURL}
  \url{https://onlinelibrary.wiley.com/doi/abs/10.1111/phpr.12221}
  \end{APACrefURL}
\newblock
\begin{APACrefDOI} \doi{https://doi.org/10.1111/phpr.12221} \end{APACrefDOI}
\PrintBackRefs{\CurrentBib}

\bibitem [\protect \citeauthoryear {%
Maudlin%
}{%
Maudlin%
}{%
{\protect \APACyear {1988}}%
}]{%
Maudlin_essence}
\APACinsertmetastar {%
Maudlin_essence}%
\begin{APACrefauthors}%
Maudlin, T.%
\end{APACrefauthors}%
\unskip\
\newblock
\APACrefYearMonthDay{1988}{}{}.
\newblock
{\BBOQ}\APACrefatitle {{The Essence of Space-Time}} {{The Essence of
  Space-Time}}.{\BBCQ}
\newblock
\APACjournalVolNumPages{PSA: Proceedings of the Biennial Meeting of the
  Philosophy of Science Association}{1988}{}{82--91}.
\newblock
\begin{APACrefURL} [{2024-03-16}]\url{http://www.jstor.org/stable/192873}
  \end{APACrefURL}
\PrintBackRefs{\CurrentBib}

\bibitem [\protect \citeauthoryear {%
Mundy%
}{%
Mundy%
}{%
{\protect \APACyear {1992}}%
}]{%
Mundy1992}
\APACinsertmetastar {%
Mundy1992}%
\begin{APACrefauthors}%
Mundy, B.%
\end{APACrefauthors}%
\unskip\
\newblock
\APACrefYearMonthDay{1992}{}{}.
\newblock
{\BBOQ}\APACrefatitle {{Space-Time and Isomorphism}} {{Space-Time and
  Isomorphism}}.{\BBCQ}
\newblock
\APACjournalVolNumPages{PSA: Proceedings of the Biennial Meeting of the
  Philosophy of Science Association}{1992}{Volume One: Contributed
  Papers}{515--527}.
\PrintBackRefs{\CurrentBib}

\bibitem [\protect \citeauthoryear {%
Norton%
}{%
Norton%
}{%
{\protect \APACyear {2019}}%
}]{%
SEP_hole}
\APACinsertmetastar {%
SEP_hole}%
\begin{APACrefauthors}%
Norton, J.%
\end{APACrefauthors}%
\unskip\
\newblock
\APACrefYearMonthDay{2019}{}{}.
\newblock
{\BBOQ}\APACrefatitle {{The Hole Argument}} {{The Hole Argument}}.{\BBCQ}
\newblock
\BIn{} E\BPBI N.~Zalta\ (\BED), \APACrefbtitle {The {Stanford} Encyclopedia of
  Philosophy} {The {Stanford} encyclopedia of philosophy}\
  (\PrintOrdinal{Summer 2019}\ \BEd).
\newblock
\APACaddressPublisher{}{Metaphysics Research Lab, Stanford University}.
\newblock
\APAChowpublished
  {\url{https://plato.stanford.edu/archives/sum2019/entries/spacetime-holearg/}}.
\PrintBackRefs{\CurrentBib}

\bibitem [\protect \citeauthoryear {%
Olver%
}{%
Olver%
}{%
{\protect \APACyear {1986}}%
}]{%
Olver_book}
\APACinsertmetastar {%
Olver_book}%
\begin{APACrefauthors}%
Olver, P.%
\end{APACrefauthors}%
\unskip\
\newblock
\APACrefYear{1986}.
\newblock
\APACrefbtitle {{Applications of Lie Groups to Differential Equations}}
  {{Applications of Lie Groups to Differential Equations}}.
\newblock
\APACaddressPublisher{}{Springer-Verlag New York}.
\PrintBackRefs{\CurrentBib}

\bibitem [\protect \citeauthoryear {%
O'Neill%
}{%
O'Neill%
}{%
{\protect \APACyear {1983}}%
}]{%
Oneill}
\APACinsertmetastar {%
Oneill}%
\begin{APACrefauthors}%
O'Neill, B.%
\end{APACrefauthors}%
\unskip\
\newblock
\APACrefYear{1983}.
\newblock
\APACrefbtitle {{Semi-Riemannian Geometry With Applications to Relativity}}
  {{Semi-Riemannian Geometry With Applications to Relativity}}.
\newblock
\APACaddressPublisher{}{Academic Press}.
\PrintBackRefs{\CurrentBib}

\bibitem [\protect \citeauthoryear {%
Pooley%
}{%
Pooley%
}{%
{\protect \APACyear {2022}}%
}]{%
Pooley_draft}
\APACinsertmetastar {%
Pooley_draft}%
\begin{APACrefauthors}%
Pooley, O.%
\end{APACrefauthors}%
\unskip\
\newblock
\APACrefYear{2022}.
\newblock
\APACrefbtitle {{The Reality of Spacetime}} {{The Reality of Spacetime}}.
\newblock
\APACaddressPublisher{}{Unpublished manuscript}.
\PrintBackRefs{\CurrentBib}

\bibitem [\protect \citeauthoryear {%
Pooley%
\ \BBA {} Read%
}{%
Pooley%
\ \BBA {} Read%
}{%
{\protect \APACyear {2022}}%
}]{%
Pooley_Read}
\APACinsertmetastar {%
Pooley_Read}%
\begin{APACrefauthors}%
Pooley, O.%
\BCBT {}\ \BBA {} Read, J.%
\end{APACrefauthors}%
\unskip\
\newblock
\APACrefYearMonthDay{2022}{}{}.
\newblock
{\BBOQ}\APACrefatitle {{On the Mathematics and Metaphysics of the Hole
  Argument}} {{On the Mathematics and Metaphysics of the Hole
  Argument}}.{\BBCQ}
\newblock
\APACjournalVolNumPages{The British Journal for the Philosophy of
  Science}{}{}{}.
\newblock
\begin{APACrefDOI} \doi{10.1086/718274} \end{APACrefDOI}
\PrintBackRefs{\CurrentBib}

\bibitem [\protect \citeauthoryear {%
Rovelli%
}{%
Rovelli%
}{%
{\protect \APACyear {2002}}%
{\protect \APACexlab {{\protect \BCnt {1}}}}}]{%
Rovelli_GPS}
\APACinsertmetastar {%
Rovelli_GPS}%
\begin{APACrefauthors}%
Rovelli, C.%
\end{APACrefauthors}%
\unskip\
\newblock
\APACrefYearMonthDay{2002{\protect \BCnt {1}}}{{\APACmonth{01}}}{}.
\newblock
{\BBOQ}\APACrefatitle {{GPS observables in general relativity}} {{GPS
  observables in general relativity}}.{\BBCQ}
\newblock
\APACjournalVolNumPages{Physical Review D}{65}{4}{}.
\newblock
\begin{APACrefURL} \url{http://dx.doi.org/10.1103/PhysRevD.65.044017}
  \end{APACrefURL}
\newblock
\begin{APACrefDOI} \doi{10.1103/physrevd.65.044017} \end{APACrefDOI}
\PrintBackRefs{\CurrentBib}

\bibitem [\protect \citeauthoryear {%
Rovelli%
}{%
Rovelli%
}{%
{\protect \APACyear {2002}}%
{\protect \APACexlab {{\protect \BCnt {2}}}}}]{%
Rovelli_partial}
\APACinsertmetastar {%
Rovelli_partial}%
\begin{APACrefauthors}%
Rovelli, C.%
\end{APACrefauthors}%
\unskip\
\newblock
\APACrefYearMonthDay{2002{\protect \BCnt {2}}}{Jun}{}.
\newblock
{\BBOQ}\APACrefatitle {Partial observables} {Partial observables}.{\BBCQ}
\newblock
\APACjournalVolNumPages{Physical Review D}{65}{12}{}.
\newblock
\begin{APACrefURL} \url{http://dx.doi.org/10.1103/PhysRevD.65.124013}
  \end{APACrefURL}
\newblock
\begin{APACrefDOI} \doi{10.1103/physrevd.65.124013} \end{APACrefDOI}
\PrintBackRefs{\CurrentBib}

\bibitem [\protect \citeauthoryear {%
Weatherall%
}{%
Weatherall%
}{%
{\protect \APACyear {2018}}%
}]{%
Weatherall_hole}
\APACinsertmetastar {%
Weatherall_hole}%
\begin{APACrefauthors}%
Weatherall, J.%
\end{APACrefauthors}%
\unskip\
\newblock
\APACrefYearMonthDay{2018}{}{}.
\newblock
{\BBOQ}\APACrefatitle {{Regarding the ‘Hole Argument’}} {{Regarding the
  ‘Hole Argument’}}.{\BBCQ}
\newblock
\APACjournalVolNumPages{The British Journal for the Philosophy of
  Science}{69}{2}{329-350}.
\newblock
\begin{APACrefURL} \url{https://doi.org/10.1093/bjps/axw012} \end{APACrefURL}
\newblock
\begin{APACrefDOI} \doi{10.1093/bjps/axw012} \end{APACrefDOI}
\PrintBackRefs{\CurrentBib}

\bibitem [\protect \citeauthoryear {%
Weatherall%
}{%
Weatherall%
}{%
{\protect \APACyear {2020}}%
}]{%
Weatherall_pre}
\APACinsertmetastar {%
Weatherall_pre}%
\begin{APACrefauthors}%
Weatherall, J.%
\end{APACrefauthors}%
\unskip\
\newblock
\APACrefYearMonthDay{2020}{}{}.
\newblock
{\BBOQ}\APACrefatitle {{Some Philosophical Prehistory of the (Earman-Norton)
  hole argument}} {{Some Philosophical Prehistory of the (Earman-Norton) hole
  argument}}.{\BBCQ}
\newblock
\APACjournalVolNumPages{Studies in History and Philosophy of Science Part B:
  Studies in History and Philosophy of Modern Physics}{70}{}{79-87}.
\newblock
\begin{APACrefURL}
  \url{https://www.sciencedirect.com/science/article/pii/S1355219820300241}
  \end{APACrefURL}
\newblock
\begin{APACrefDOI} \doi{https://doi.org/10.1016/j.shpsb.2020.02.002}
  \end{APACrefDOI}
\PrintBackRefs{\CurrentBib}

\end{thebibliography}

\end{document}